\documentclass[a4paper,12pt]{article}

\usepackage{psfrag}
\usepackage{mathrsfs}

\usepackage{graphicx}
\usepackage[margin=15pt,font=small,labelfont=bf,labelsep=period]{caption}
\usepackage{amsmath}
\usepackage{amssymb}
\usepackage{amsfonts}
\usepackage{mathtools}
\usepackage{bm}
\usepackage{dsfont}
\usepackage{cite}
\usepackage{setspace}
\usepackage{environ,letltxmacro}
\usepackage{tikz}
\usetikzlibrary{arrows,decorations.pathreplacing}
\usepackage{enumitem}
\usepackage{hyperref}
\usepackage{relsize}
\usepackage[a4paper,textwidth=16.04cm,textheight=22cm,footskip=15mm]{geometry}

\usepackage{empheq}
\usepackage{environ}
\NewEnviron{boxalign}{\begin{empheq}[box=\fbox]{align} \BODY \end{empheq}}

\numberwithin{equation}{section}

\LetLtxMacro\oldequation\equation
\LetLtxMacro\endoldequation\endequation
\let\equation\relax
\let\endequation\relax
\NewEnviron{equation}[1][\empty]{%
 \ifx \empty#1
  \oldequation \BODY \endoldequation
 \else
  \ifx b#1
   \oldequation \fbox{$\displaystyle \BODY $} \endoldequation
  \else
   \oldequation \BODY \endoldequation
  \fi
 \fi}

\begin{document}

\begin{titlepage}

\title{
\begin{flushright}
\normalsize{MITP/16-121}
\bigskip
\vspace{1cm}
\end{flushright}
Composite Operators in Asymptotic Safety\\[2mm]
}

\date{}

\author{Carlo Pagani and Martin Reuter\\[3mm]
{\small Institute of Physics, PRISMA \& MITP,}\\[-0.2em]
{\small Johannes Gutenberg University Mainz,}\\[-0.2em]
{\small Staudingerweg 7, D--55099 Mainz, Germany}
}

\maketitle
\thispagestyle{empty}

\vspace{2mm}
\begin{abstract}

We study the role of composite operators in the Asymptotic Safety program for quantum gravity.
By including in the effective average action an explicit dependence on new sources we are able to keep track of
operators which do not belong to the exact theory space and/or are normally discarded in a truncation.
Typical examples are geometric operators such as volumes, lengths, or geodesic distances.
We show that this set-up allows to investigate the scaling properties of various interesting
operators via a suitable exact renormalization group equation. 
We test our framework in several settings, including Quantum Einstein Gravity,
the conformally reduced Einstein-Hilbert truncation, and two dimensional quantum gravity.
Finally, we briefly argue that our construction paves the way to approach 
observables in the Asymptotic Safety program.

\end{abstract}

\end{titlepage}

\newpage

\begin{spacing}{1.1}


\section{Introduction}

The construction of a well defined path integral for quantum gravity
is at the heart of the Asymptotic Safety (AS) program \cite{W80,R98}. Ultimately, however,
the construction of this path integral should allow the evaluation
of observable quantities, typically expectation values of composite operators. 
In this work we make a first step towards
the latter goal. 
So far the focus of most of the investigations regarding
AS has been devoted to probing the existence of a suitable UV fixed point. 
The framework employed in these investigations involves
the Effective Average Action (EAA) and its Renormalization Group (RG) \cite{W93}.
In this setting, truncations of increasing complexity have been analyzed
including bimetric ans{\"a}tze, higher derivative terms and infinite dimensional truncations,
 see~\cite{Codello:2013fpa,Becker:2014qya,Demmel:2015oqa,Christiansen:2015rva,Gies:2016con,Ohta:2015fcu,Ohta:2015zwa,Dietz:2016gzg}
for some of the most recent works.

Our aim in the present work is different. Let us suppose that we have a reasonably
well approximated gravitational EAA, can we extract all possible information
from the EAA alone? We wish to argue that this is not always the case,
there are instances in which further efforts are required. 

Before entering into the technical aspects, we would like to recall the special status
of quantum gravitational theories with respect to non-gravitational
ones. In particular, observables in a quantum gravitational theory
are required to be diffeomorphism invariant, i.e.~gauge invariant.
In turn this implies that we cannot think of an observable as depending
on a point of the spacetime manifold, since a diffeomorphism transformation
would change it. Instead, one may consider quantities integrated over
all spacetime. However, such quantities are rather distant from our
intuition, which is trained to think in terms of ``localized'' quantities. 

A possible way out of this conceptual dilemma is to recall that
in performing a measurement we actually check for the coincidence
of events, like for instance that a photon hits our experimental apparatus. 
The fact that the photon hits the detector is invariant under diffeomorphisms
since the statement that photon and detector are at the
same spacetime point remains true after a diffeomorphism transformation
is applied.
 
To implement a consistent description of the system
plus the apparatus in the field theoretic language is not an easy
task. Following DeWitt \cite{DeWitt:1962cg} one may modify the action
functional via $S\rightarrow S+\varepsilon A$, where the last term
describes the coupling of the system to the detector.
As a result we observe that, purely at the quantum field theoretic
level, information regarding the new operator $A$ is required.
However, in general this information is not encoded automatically in the EAA. 

To properly define observables in quantum gravity also other approaches
have been considered. For instance one may use scalar fields to localize
observables or define correlation functions at fixed geodesic distance,
we refer the reader to \cite{DeWitt:1962cg,Rovelli:1990ph,Giddings:2005id} more details.

With regard to the gravitational EAA formalism, all these approaches have a common
feature: they require information about operators which usually are not taken
into account in a truncated EAA, at any realistic level of complexity. 
For instance, it is hard to imagine a truncation for the gravitational EAA to contain information on
the geodesic distance of two given points on the spacetime manifolds,
a quantity that appears in many observables of practical interest, however, \cite{Ambjorn:1997di,Hamber:2009zz}.

In order to obtain information regarding an arbitrary operator in
a quantum field theoretic framework one can couple it to an external source so that 
it can be inserted into correlation
functions by taking functional derivatives with
respect to the source. This formalism goes under the name of the 
{\it composite operator formalism}.
It allows to define, and actually compute correlation functions of
not only elementary fields, but also of more complicated local operators
at a given spacetime point. The main task of this work is to investigate
the composite operator formalism and its application within the framework
of the gravitational EAA.

The introduction of composite operators is unavoidable also in many other
cases. For example, let us consider the correlation function between metric operators at different points
in the vielbein formalism. In this case the metric itself is a composite
operator which can be meaningfully defined only via a suitable regularization and renormalization procedure
over and above the usual renormalization of couplings in the EAA.

In the present work we are going to provide the basic framework to properly define this
type of operators in the EAA formalism, and we consider some explicit
examples that occur in the AS context.

This paper is organized as follows. In section \ref{sec:Floating-normalization-point-and-FRG}
we revisit an argument which allows to define the scaling dimensions
of operators straightforwardly in the EAA framework. In section \ref{sub:Composite-operators-in-FRG}
we include composite operators into the EAA by coupling them to an external
source, discuss possible approximations, and show how to compute the scaling properties
of these composite operators. 

In section \ref{sec:Composite-metrics-in-the-CREH}
we consider the conformally reduced Einstein-Hilbert (CREH) truncation,
a simple model which mimics many features of Quantum Einstein Gravity (QEG).
In this setting the metric is parametrized by a dynamical conformal
factor times a fixed reference metric. The conformal factor is actually
a composite operator of the elementary quantum field, and so the metric
in the CREH setting can be thought of as a toy model for the composite
metric of the vielbein formalism. We discuss the definition of the
metric as a composite operator in this framework. 

In section \ref{sec:Volume-operators-in-AS}
we investigate the scaling properties of two geometrical objects within QEG: the volume, and
the length of curves. 

Finally, in section \ref{sec:Two-dimensional-QG},
we study various composite operators in two dimensional quantum gravity. 
The two dimensional case is interesting for various
reasons. First, there is a variety of results coming from other approaches
and techniques, such as conformal field theory, to which one may compare
the findings given by our framework. Second, two dimensional Asymptotic
Safety has been recently investigated in detail \cite{Nink:2015lmq} and,
among other things, it has been possible to test the compatibility
between the presence of a non-Gaussian fixed point and unitarity in
this context. Thus, it is natural to ask what kind of consequences
such a fixed point bears for geometrical objects like the volume operator
or the length of a curve.
Furthermore, in the Appendix we show as an example how our approach leads to the 
familiar KPZ scaling relations for gravitationally dressed operators.


\section{Scaling arguments and functional renormalization group \label{sec:Floating-normalization-point-and-FRG}}

The Effective Average Action (EAA) is
a scale dependent generalization of the standard effective action \cite{W93}. 
One introduces a scale $k$ below which the integration
of momentum modes is suppressed. This is achieved by adding the cutoff term
$\Delta S_{k}=\frac{1}{2}\int\chi {\cal R}_{k}\chi$ to the bare action,
with ${\cal R}_{k}$ being a suitable kernel. The scale dependence of the effective
average action satisfies the following exact functional RG equation or ``FRGE'' \cite{W93}
\begin{eqnarray}
\partial_{t}\Gamma_{k} & = & \frac{1}{2}\mbox{Tr}\left[\left(\Gamma_{k}^{\left(2\right)}+{\cal R}_{k}\right)^{-1}\partial_{t}{\cal R}_{k}\right]\label{eq:flow_eq_EAA}
\end{eqnarray}
where $\Gamma_{k}^{\left(2\right)}$ is the Hessian of the effective
average action, $\Gamma_{k}$, and $t \equiv \log k$. This equation can be concretely employed
after implementing some approximation scheme. 

In this section we briefly review an argument which allows us to deduce
the scaling properties of any operator in the EAA formalism \cite{Pagani:2016pad}. 
First let us note that to uniquely solve the flow equation (\ref{eq:flow_eq_EAA})
a boundary condition must be given. Such a boundary condition is imposed
at a certain scale $\mu$, which we call the {\it floating normalization point}.
The dependence of the EAA on the scale $\mu$ has been studied in
detail in \cite{Pagani:2016pad}, and in this section we shall revisit the dependence
in the framework of the gravitational EAA. In particular, we shall
see that the EAA is invariant under suitable changes of the boundary
condition. Such invariance properties allow one to write down an equation
fully analogous to the Callan-Symanzik equation of standard quantum field
theory. This equation, together with simple dimensional analysis,
allows one to discuss the scaling properties of the theory straightforwardly.

Let us consider a theory space parametrized by a set of dimensionless
couplings $\left\{ \tilde{g}_{i}\right\} $. The RG flow is described
by a system of differential equations:
\begin{eqnarray}
\partial_{t}\tilde{g}_{i} & = & f_{i}\left(\left\{ \tilde{g}_{j}\right\} \right)\,,\label{eq:RG_flow_coupling}
\end{eqnarray}
to which one associates boundary conditions like\footnote{
For lack of a better word we refer to it as a ``boundary'' condition even if $\mu$ is an inner point
of the $k$-interval under consideration.}
\begin{eqnarray}
\tilde{g}_{i}\left(\mu\right) & = & \tilde{g}_{i}^{\left({\rm R}\right)}\,,\label{eq:boundary_cond_g_i_tilde}
\end{eqnarray}
where the ``renormalized'' couplings $\tilde{g}_{i}^{\left({\rm R}\right)}$ are given numbers. 
By imposing a boundary condition we select a specific trajectory on theory
space. Let us denote this solution by $\tilde{g}_{i}^{\left({\rm sol}\right)}\left(k;\mu,\tilde{g}_{i}^{\left({\rm R}\right)}\right)$,
where we made explicit its dependence on the boundary values $\tilde{g}_{i}^{\left({\rm R}\right)}$
and the scale $\mu$. 

Clearly, if one chooses another set of boundary values which however still correspond
to some point along this trajectory then the solution of the flow equation will be the
very same trajectory again. 
To cast this simple fact into a mathematical formula let
us consider the specific solution of equation (\ref{eq:RG_flow_coupling})
associated with the boundary condition (\ref{eq:boundary_cond_g_i_tilde}).
Now we want to change the boundary condition (\ref{eq:boundary_cond_g_i_tilde})
to an equivalent boundary condition along the trajectory, i.e.~we
move $\mu$ to some other scale $\mu^{\prime}$ and change the couplings
accordingly. This is achieved by infinitesimally translating $\mu\rightarrow\mu^{\prime}=\mu+\varepsilon$
and $\tilde{g}_{i}^{\left({\rm R}\right)}=\tilde{g}_{i}\left(\mu\right)\rightarrow\tilde{g}_{i}^{\left(R\right)\prime}=\tilde{g}_{i}\left(\mu^{\prime}\right)=\tilde{g}_{i}\left(\mu\right)+\varepsilon\partial_{\mu}\tilde{g}_{i}\left(\mu\right)$.
The fact that these two boundary conditions are associated to the same
solution implies that:
\begin{eqnarray*}
\tilde{g}_{i}^{\left({\rm sol}\right)}\left(k;\mu,\tilde{g}_{i}^{\left({\rm R}\right)}\right) & = & \tilde{g}_{i}^{\left({\rm sol}\right)}\left(k;\mu^{\prime},\tilde{g}_{i}^{\left(R\right)\prime}\right)\\
 & \cong & \tilde{g}_{i}^{\left({\rm sol}\right)}\left(k;\mu,\tilde{g}_{i}^{\left({\rm R}\right)}\right)+\varepsilon\left(\partial_{\mu}+\partial_{\mu}\tilde{g}_{j}\left(\mu\right)\frac{\partial}{\partial\tilde{g}_{j}^{\left(R\right)}}\right)\tilde{g}_{i}^{\left({\rm sol}\right)}\left(k;\mu,\tilde{g}_{i}^{\left({\rm R}\right)}\right) \,.
\end{eqnarray*}
As a consequence, it follows that
\begin{equation}
 \fbox{$\displaystyle
\left(\mu\partial_{\mu}+\beta_{j}\frac{\partial}{\partial\tilde{g}_{j}^{\left(R\right)}}\right)\tilde{g}_{i}^{\left({\rm sol}\right)}\left(k;\mu,\tilde{g}_{i}^{\left({\rm R}\right)}\right)  =  0\,,
  $}
\end{equation}
where $\beta_{j}\equiv\beta_{j}\left(\tilde{g}_{i}^{\left({\rm R}\right)}\right)$.
The same reasoning straightforwardly applies to the entire EAA. 
Thereby a wave function renormalization $Z_{k}$ can be conveniently introduced considering
ans{\"a}tze of the following form:\footnote{Since the total number of running couplings (including the wave function
renormalization constant) should be $n$ we set $g_{i}=1$ for some
$i$. For instance, in the case of a scalar field theory one usually
chooses to write the kinetic term $\frac{1}{2}Z_{k}\left(\partial\varphi\right)^{2}$,
with no coupling $g_{i}$ in front.}
\begin{eqnarray*}
\Gamma_{k}\left[\varphi\right] & = & \sum_{i=1}^{n}g_{i}O_{i}\left(Z_{k}^{1/2}\varphi \right)\,.
\end{eqnarray*}
Here we made explicit the inessential nature of $Z_{k}$. The anomalous
dimension of the elementary field $\varphi$ corresponds to $\eta\equiv -Z_{k}^{-1}\partial_{t}Z_{k}$, 
see \cite{Pagani:2016pad} for a detailed discussion. One
thus has an equation which is fully similar to the {\it Callan-Symanzik equation}:
\begin{equation}
 \fbox{$\displaystyle
\left(\mu\partial_{\mu}+\beta_{j}\frac{\partial}{\partial\tilde{g}_{j}^{\left(R\right)}}-\eta\varphi\cdot\frac{\delta}{\delta\varphi}\right)\Gamma_{k}\left[\varphi\right]  =  0\,.\label{eq:Callan_Symanzik_for_EAA}
  $}
\end{equation}

Equation (\ref{eq:Callan_Symanzik_for_EAA}) can be used to deduce
scaling properties of correlation functions at a fixed point. 
To do so one considers equation (\ref{eq:Callan_Symanzik_for_EAA}) together with
an Euler-type differential equation (homogeneity relation) which stems from
dimensional analysis. 

As an example, let us consider the propagator
of a scalar field with mass dimension $\left[\varphi\right]=\frac{d-2}{2}$.
In the fixed point regime, with $\Gamma = \Gamma_{k\rightarrow 0}$, we obtain from equation (\ref{eq:Callan_Symanzik_for_EAA})
and dimensional analysis (see \cite{Pagani:2016pad} for details):
\begin{equation}
\left\{ \begin{array}{rr}
\left[ \mu\partial_{\mu}-\eta\right]\Gamma^{\left(2\right)} & =0\\
\left[ \mu\partial_{\mu}+p\partial_{p}+\left(d-2\right)\right]\Gamma^{\left(2\right)} & =0
\end{array}\right.\,.\label{eq:scalar-fixed-point-propagator}
\end{equation}
Now we eliminate the $\mu\partial_\mu$ term from these two equations. Moreover, taking
into account the overall delta function entailing momentum conservation
in $\Gamma^{\left(2\right)}$ and factoring it out, one obtains
\begin{equation}
\left[ p\partial_{p}-\left(2-\eta\right)\right] \Gamma^{\left(2\right)}=0\,,  \label{eq:5prime-martin-corrections}
\end{equation}
where we denoted $\Gamma^{\left(2\right)}$ the two-point function
with the delta function stripped away. Remarkably we note that equation
(\ref{eq:5prime-martin-corrections}) just derives from the (here assumed) existence of a fixed
point. Different fixed point propagators are distinguished by the
different values of the anomalous dimension. 

One can repeat the same logic for the graviton propagator. Let us
remark that in the case of gravitational theories one can either consider
the coordinates dimensionful and the metric dimensionless or vice-versa.
Either ways, the above reasoning leads straightforwardly to a propagator
of the type $p^{4}$ as it has been already noted in \cite{Lauscher:2001ya,Lauscher:2002sq,Lauscher:2005qz}.
Such propagator can be viewed as a two dimensional propagator hinting
to a dimensional reduction phenomenon \cite{NR06,Lauscher:2005qz,Reuter:2011ah}. The computation
of the anomalous dimension in the spirit mentioned above (leading
to a propagator of the type $p^{4-\eta}$) has been performed in very
few truncations, see for instance \cite{Codello:2013fpa,Becker:2014qya,Christiansen:2014raa}. 

As far as composite operators are concerned, the argument outlined in this section can be straightforwardly generalized 
and allows to identify their scaling dimensions. In section \ref{sub:Composite-operators-in-FRG}, 
we shall define the scaling dimension of composite operators and see how they can be estimated.

\section{Composite operators in the functional renormalization group and Asymptotic
Safety \label{sub:Composite-operators-in-FRG}}

In the functional integral formulation of standard quantum field theory one deals with composite operators
by coupling them to external sources so that one obtains insertions of composite
operators in correlation functions by taking suitable functional derivatives
of the path integral \cite{Itzykson:1980rh}. In the Effective Average Action
formalism this step is not often made, one of the reasons being that frequently one is interested
in the properties of a system at a fixed point, which one describes
by the critical exponents associated to the couplings $\left\lbrace g_i^* \right\rbrace $ of 
the operators present in the truncation, $O_i$. However, there
are several situations in which one may wish to couple some operators
to their respective sources and carry out the associated renormalization
procedure.

First of all, in order to solve equation (\ref{eq:flow_eq_EAA}),
a truncation ansatz is typically used. 
This is one possible reason why not ``all'' operators
are present in the ansatz for the EAA. If one was interested in the
scaling of an operator $O$ which, for any reason, is not present
in the truncation ansatz for the EAA, a procedure analogous to the one adopted
in standard quantum field theory is very helpful and gives a first
estimate of the scaling properties of the operator. 

Moreover, there are operators which one is not able to treat directly even in a full
fledged EAA. As an example, let us consider the metric in the vielbein
formalism, i.e.~$g_{\mu\nu}=e_{\mu}^{a}e_{\nu}^{b}\eta_{ab}$. 
If $e^a_\mu$ is taken to be an elementary field (possibly together with the spin-connection
as in the Riemann-Cartan theory) then $g_{\mu\nu}\left(x\right)$ is neither an elementary field,
it is an operator product of two fields, nor it is an invariant built from elementary fields, 
i.e.~it is not contained in the exact theory space even.
In this case the metric is a composite operator of spin two, 
and in order to define meaningful correlation functions of the
metric one needs to regularize and renormalize the operator $g_{\mu\nu}$. 
This begins with coupling $g_{\mu\nu}$ to a spin two source. Similar
considerations hold for many other interesting operators as we shall see
later on.

Let us review how one can deal with composite operators in the functional
renormalization framework. We denote $\varepsilon\left(x\right)$
the source and consider the expectation value\footnote{Whenever a dot appears in a mathematical expression, e.g.~$f\cdot g$,
the DeWitt condensed notation is used, meaning that integration and
index summation is intended.} 
\begin{eqnarray*}
\langle O\left( x \right) \rangle & = & {\cal N}\int{\cal D}\chi\,O\left( x \right) e^{-S}\\
 & = & -\frac{\delta}{\delta\varepsilon \left( x \right)}{\cal N}\int{\cal D}\chi\,e^{-S-\varepsilon\cdot O}\Bigr|_{\varepsilon=0}\,,
\end{eqnarray*}
where ${\cal N}$ is a suitable normalization constant. Then we define the
generating functional $W\left[J,\varepsilon\right]$ for the connected
Green's functions associated to the modified action $S+\varepsilon\cdot O$:
\begin{eqnarray*}
e^{W\left[J,\varepsilon\right]} & \equiv & \int{\cal D}\chi\,e^{-S-\varepsilon\cdot O+J\cdot\chi}\,.
\end{eqnarray*}
The associated effective action is obtained via a Legendre transform
\begin{eqnarray*}
\Gamma\left[\varphi,\varepsilon\right] & = & J\cdot\varphi-W\left[J,\varepsilon\right]\,,\;\varphi= \frac{\delta W}{\delta {J}} \,.
\end{eqnarray*}
It is straightforward to check that
\begin{eqnarray*}
\frac{\delta\Gamma}{\delta\varepsilon}\left[\varphi,\varepsilon\right] & = & -\frac{\delta W}{\delta\varepsilon}\left[J,\varepsilon\right]\,,
\end{eqnarray*}
which tells us that we can extract the renormalization regarding a
single insertion of a composite operator directly considering a single
functional derivative with respect to $\varepsilon\left(x\right)$
of the EAA. One can repeat the derivation of the
FRGE in the case of $\Gamma_k \left[ \varphi, \varepsilon \right]$.
From its $\varepsilon$-derivative we find the following exact flow equation 
associated to the composite operator \cite{Igarashi:2009tj,Pawlowski:2005xe,Pagani:2016pad}:
\begin{eqnarray*}
\partial_{t}\left(\frac{\delta}{\delta\varepsilon}\Gamma_{k}\left[\varphi,\varepsilon\right]\right)\Bigr|_{\varepsilon=0} & = & -\frac{1}{2}\mbox{Tr}\left[\left(\Gamma_{k}^{\left(2\right)}+{\cal R}_{k}\right)^{-1}\frac{\delta\Gamma_{k}^{\left(2\right)}}{\delta\varepsilon}\left(\Gamma_{k}^{\left(2\right)}+{\cal R}_{k}\right)^{-1}\partial_{t}{\cal R}_{k}\right]\Bigr|_{\varepsilon=0}\,.
\end{eqnarray*}
We can avoid performing the functional derivative with respect to
$\varepsilon$ and just compare order by order in $\varepsilon$.
Clearly, since we are interested just in a single insertion of the
composite operator, we can limit ourselves to consider the case where
$\varepsilon^{2}=0$. Furthermore we denote
\begin{eqnarray*}
\left[O_{k}\right]_{i} & \equiv & \frac{\delta}{\delta\varepsilon_{i}}\Gamma_{k}\left[\varphi,\varepsilon_{j}\right]
\end{eqnarray*}
where $k$ indicates the RG scale and the subscript $i$ labels $n$
different composite operators. We can rewrite the flow equation for
composite operators as \cite{Pagani:2016pad}
\begin{equation}
 \fbox{$\displaystyle
\partial_{t}\left(\varepsilon\cdot\left[O_{k}\right]\right) = -\frac{1}{2}\mbox{Tr}\left[\left(\Gamma_{k}^{\left(2\right)}+{\cal R}_{k}\right)^{-1}\left(\varepsilon\cdot\left[O_{k}\right]^{\left(2\right)}\right)\left(\Gamma_{k}^{\left(2\right)}+{\cal R}_{k}\right)^{-1}\partial_{t}{\cal R}_{k}\right]\,.\label{eq:flow_eq_composite_operator_epsilon}
  $}
\end{equation}

To concretely solve equation (\ref{eq:flow_eq_composite_operator_epsilon})
some approximation must be implemented. In particular one may expand
the composite operator $\left[O_{k}\right]_{i}$ in a basis of $k$-independent operators
$\left\lbrace O_{i},\,i=1,\cdots,n \right\rbrace$. In this case
\begin{eqnarray}
\left[O_{k}\right]_{i} & = & \sum_{j=1}^{n}Z_{ij} \left( k \right)  O_{j}\,.\label{eq:ansatz_composite_oper_Z_Oi}
\end{eqnarray}
By following the reasoning of
section \ref{sec:Floating-normalization-point-and-FRG} one can show
that the scaling operators of the theory have dimensions, quantum corrections
included, given by the eigenvalues of the matrix
\begin{equation}
d_{i}\delta_{ij}+\left(Z^{-1}\partial_{t}Z\right)_{ij}\,,
\end{equation}
where $d_{i}$ is the (classical) mass dimension of the operator $O_{i}$ \cite{Pagani:2016pad}. 

The crucial matrix $\gamma_{Z,ij}\equiv\left(Z^{-1}\partial_{t}Z\right)_{ij}$
can be directly found manipulating equation (\ref{eq:flow_eq_composite_operator_epsilon}).
Inserting the ansatz (\ref{eq:ansatz_composite_oper_Z_Oi}), and taking
a functional derivative with respect to $\varepsilon_{i}$, we find
\begin{eqnarray*}
\sum_{j}\partial_{t}\left(Z_{ij}O_{j}\left(x\right)\right) & = & -\frac{1}{2}\mbox{Tr}\left[\left(\Gamma_{k}^{\left(2\right)}+{\cal R}_{k}\right)^{-1}\left(\sum_{j}Z_{ij}O_{j}^{\left(2\right)}\left(x\right)\right)\left(\Gamma_{k}^{\left(2\right)}+{\cal R}_{k}\right)^{-1}\partial_{t}{\cal R}_{k}\right]\,,
\end{eqnarray*}
which implies the final result, for the general case with operator mixing,
\begin{eqnarray}
\sum_{j}\gamma_{Z,ij}O_{j}\left(x\right) & = & -\frac{1}{2}\mbox{Tr}\left[\left(\Gamma_{k}^{\left(2\right)}+{\cal R}_{k}\right)^{-1}\left(O_{i}^{\left(2\right)}\left(x\right)\right)\left(\Gamma_{k}^{\left(2\right)}+{\cal R}_{k}\right)^{-1}\partial_{t}{\cal R}_{k}\right]\,.\label{eq:flow_eq_gamma_Z}
\end{eqnarray}

In the present work we shall mainly limit ourselves to non-mixing
ans\"{a}tze for the composite operators. This means that we shall
consider composite operators approximated by the simple parametrization
$\left[O_{k}\right]=Z_{O} \left(k \right)O$. Such an operator aquires an anomalous
dimension given by $Z_{O}^{-1}\partial_{t}Z_{O}$ which can be read
off from 
\begin{equation}
 \fbox{$\displaystyle
\gamma_{Z_{O}}O\left(x\right) =  -\frac{1}{2}\mbox{Tr}\left[\left(\Gamma_{k}^{\left(2\right)}+{\cal R}_{k}\right)^{-1}\left(O^{\left(2\right)}\left(x\right)\right)\left(\Gamma_{k}^{\left(2\right)}+{\cal R}_{k}\right)^{-1}\partial_{t}{\cal R}_{k}\right]\,.\label{eq:flow_eq_gamma_Z_no_mixing}
  $}
\end{equation}

For the sake of comparison with other results in the literature, it
is useful to work out the relation between scaling operators defined
by means of explicit introduction of the sources and those found by
{\it linearizing the RG flow around the fixed point}. Let us consider an
ansatz for the EAA expanded in the basis of operators $O_{i}$:
\begin{eqnarray*}
\Gamma_{k} & = & \sum_{i=1}^{n}g_{i} \left(k \right) O_{i}\,.
\end{eqnarray*}
Here we consider the basis of operators $O_{i}$ to be the same that
we used previously for the composite operators. Under these approximations
it is straightforward to conclude from the flow equation that 
\begin{eqnarray*}
\sum_{j=1}^{n}\beta_{j}O_{j} & = & \frac{1}{2}\mbox{Tr}\left[\left(\sum_{j=1}^{n}g_{j}O_{j}^{\left(2\right)}+{\cal R}_{k}\right)^{-1}\partial_{t}{\cal R}_{k}\right]\,.
\end{eqnarray*}
Taking a derivative with respect to the coupling $g_{i}$ we find
\begin{eqnarray}
\sum_{j=1}^{n}\partial_{g_{i}}\beta_{j}O_{j} & = & -\frac{1}{2}\mbox{Tr}\left[\left(\sum_{j=1}^{n}g_{j}O_{j}^{\left(2\right)}+{\cal R}_{k}\right)^{-1}O_{i}^{\left(2\right)}\left(\sum_{j=1}^{n}g_{j}O_{j}^{\left(2\right)}+{\cal R}_{k}\right)^{-1}\partial_{t}{\cal R}_{k}\right].\quad\quad\label{eq:flow_eq_partial_gi_betaj}
\end{eqnarray}
Comparing (\ref{eq:flow_eq_gamma_Z}) with (\ref{eq:flow_eq_partial_gi_betaj})
we conclude that, at the dimensionful level,
\begin{eqnarray*}
\partial_{g_{i}}\beta_{j} & = & \gamma_{Z,ij}\,,
\end{eqnarray*}
which can be rewritten in terms of dimensionless couplings $\tilde{g}_{j}$
as
\begin{eqnarray}
K_{ia} \left(d\delta_{ab}+\partial_{\tilde{g}_{a}}\tilde{\beta}_{b} \right) K_{bj}^{-1}
& = & d_{i}\delta_{ij}+\gamma_{Z,ij}\,,\label{eq:connection_critical_exp_anomalous_dim}
\end{eqnarray}
where $d$ is the spacetime dimension and $K_{ij}\equiv k^{d_i} \delta_{ij}$. 
Thus, under these approximations,
{\it the scaling dimensions found by diagonalizing the matrix $d_{i}\delta_{ij}+\gamma_{Z,ij}$
are exactly the same as those found by linearizing the RG flow and diagonalizing $d\delta_{ij}+\partial_{\tilde{g}_{i}}\tilde{\beta}_{j}$}.
(Recall also that the negative eigenvalues of $\partial_{\tilde{g}_{i}}\tilde{\beta}_{j}$ are the fixed point's critical exponents $\theta_i$.) 

The usefulness of adopting the composite operator point of view is
that there may be cases in which some operators are not included in
a truncation but one would like to have information about their renormalization
and scaling properties. 
As far as the Asymptotic Safety scenario is
concerned, an interesting example is given by the scaling properties
of the length of curves, and geodesics in particular, which usually are not considered
as a part of the EAA. 
Of course, in order to explore gravitational observables, further efforts are
required since one needs to identify suitable diffeomorphism invariant
operators. Possibly, this can be achieved by having at our disposal further
fields which allow to ``localize'' quantities in spite of an overall
integration over the manifold, see \cite{DeWitt:1962cg,Rovelli:1990ph} for a detailed
description. In this work we shall not pursue this approach further but simply consider the renormalization of possibly interesting
composite operators.

Finally, we note that scaling properties of correlation functions
involving certain suitable composite operators are also essential in order to compare
different approaches to two dimensional quantum gravity \cite{KPZ88,DDK88}.
Possibly, one may find similar comparisons between four dimensional
Asymptotic Safety and other approaches to $4D$ quantum gravity, like CDT, for example. 
This is a further motivation for the present investigation.

\section{Composite metrics in the CREH truncation \label{sec:Composite-metrics-in-the-CREH}}

In this section we consider the conformally reduced Einstein-Hilbert
(CREH) truncation and evaluate the scaling properties of various 
operators in this setting. Interestingly, in the CREH truncation the
metric is a composite operator and therefore this framework constitutes
an instructive toy model to see which types of computations are required
in more refined cases, such as the composite metric in the vielbein
formalism. 
In section \ref{sub:The-CREH-action} we
briefly recall the CREH truncation, and in section \ref{sub:Composite-metrics}
we treat the composite metric operator in the CREH truncation by means
of two different approaches that will turn out equivalent in the end.

\subsection{The CREH action \label{sub:The-CREH-action}}

The CREH truncation is inspired by the classical action functional:
\begin{eqnarray}
S\left[g_{\mu\nu}\right] & = & \frac{1}{16\pi G}\int d^{d}x\,\sqrt{g}\left(2\Lambda-R\right)\label{eq:EH_truncation}
\end{eqnarray}
evaluated for arguments $g_{\mu\nu}$ which are given by a dynamical
conformal factor times a fixed reference metric $\hat{g}_{\mu\nu}$:
\begin{eqnarray}
g_{\mu\nu} & = & \phi^{2\nu\left(d\right)}\hat{g}_{\mu\nu}\,.\label{eq:conformally_reduced_metric_1}
\end{eqnarray}
The conformal factor is written as a power
of the elementary dynamical field, $\phi$, the choice for the exponent being 
\begin{eqnarray*}
\nu\left(d\right) & \equiv & \frac{2}{d-2}\,.
\end{eqnarray*}
The exponent $2\nu$ is integer only in the special dimensions $d=3$,
$d=4$ and $d=6$, respectively. (See table \ref{table:Conf_Factors_Vol_Operators}.)
The distinguished parametrization of the conformal factor in (\ref{eq:conformally_reduced_metric_1}) has the
``miraculous'' property that, with this choice, the restricted Einstein-Hilbert
action $S\left[\phi\right]\equiv S\left[\phi^{2\nu}\hat{g}\right]$
has a standard quadratic kinetic term for $\phi$. The only self-interactions
of $\phi$ are due to the cosmological constant then. 

Furthermore allowing the cosmological and the Newton constants in $S\left[\phi\right]$
to be scale dependent ($\Lambda\rightarrow\Lambda_{k},G\rightarrow G_{k}$),
this functional reads
\begin{eqnarray}
\Gamma_{k}\left[\phi\right] & = & -\frac{1}{8\pi\xi\left(d\right)G_{k}}\int d^{d}x\sqrt{\hat{g}}\left(\frac{1}{2}\hat{g}^{\mu\nu}\partial_{\mu}\phi\partial_{\nu}\phi+\frac{1}{2}\xi\left(d\right)\hat{R}\phi^{2}-\xi\left(d\right)\Lambda_{k}\phi^{\frac{2d}{d-2}}\right),\;\;\;\label{eq:CREH_truncation_generic_dimension}
\end{eqnarray}
with $\hat{R}$ the curvature scalar of $\hat{g}_{\mu\nu}$, and 
\begin{eqnarray*}
\xi\left(d\right) & \equiv & \frac{d-2}{4\left(d-1\right)}\,.
\end{eqnarray*}
We shall refer to the action (\ref{eq:CREH_truncation_generic_dimension})
as to the CREH ansatz the for EAA of conformally reduced gravity. 

Despite its simplicity, this model captures many features of full fledged truncations
in Quantum Einstein Gravity (QEG) with all the modes of the metric retained.
In particular the RG flow is qualitatively identical to that of full
QEG, displaying in particular a non-trivial fixed point (NGFP). It
has been studied in detail in 
\cite{Reuter:2008wj,Reuter:2008qx,Manrique:2009uh,Manrique:2009tj,Machado:2009ph,Dietz:2015owa,Labus:2016lkh,Dietz:2016gzg}.

Note that when the cosmological constant is negligible ($\Lambda_k =0$) and correspondingly we choose
a flat background ($\hat{g}_{\mu\nu}= \delta_{\mu\nu}$, $\hat{R}=0$) the CREH action (\ref{eq:CREH_truncation_generic_dimension})
reduces to $\Gamma_k \propto \int \left( \partial_\mu \phi \right)^2$.
So one could think that we are dealing ``only with a free theory'' which has no interesting
renormalization behaviour.
But clearly this is false: In the model at hand even the most basic operator of physical interest, namely $g_{\mu\nu}$,
is a non-trivial composite operator of the elementary quantum field, $\phi$.
Hence there is a large class of physically relevant renormalization effects, 
namely those related to operator products,
which are not reflected by the running of the EAA in any way!

\subsection{Composite metric operators \label{sub:Composite-metrics} }

When using the parametrization (\ref{eq:conformally_reduced_metric_1})
the metric $g_{\mu\nu}$ becomes proportional to a power of the dynamical,
i.e.~quantum, field $\phi$. Thus the metric $g_{\mu\nu}$ is a composite
operator\footnote{
The case $d=6$ where $g_{\mu\nu}=\phi \hat{g}_{\mu\nu}$ happens to be linear in the quantum field and is special,
see \cite{Nink:2015lmq} for a discussion on this point.} 
which must be dealt with by a suitable renormalization procedure. 

To see why such a renormalization is needed let us consider $d=4$
dimensions where we have $g_{\mu\nu}=\phi^{2}\hat{g}_{\mu\nu}$. This
poses the problem of defining the composite operator $\phi^{2}$.
It is intructive to consider explicitly the correlation function $\langle\phi\left(x\right)\phi\left(y\right)\rangle$
in the EAA formalism and to explore how this two-point function becomes
ill-defined in the limit $y\rightarrow x$. 
To properly define $\lim_{x\rightarrow y}\langle\phi\left(x\right)\phi\left(y\right)\rangle$,
we shall need a further regularization scheme, this time for the UV, besides the mode
suppression built into the EAA. The pertinent \break
(re-)normalization procedure
will yield a meaningful the composite operator $\phi^{2}$ then.

Regarding the connected two-point function $\langle\phi\left(x\right)\phi\left(y\right)\rangle$,
in the EAA formalism it is most conveniently obtained from the inverse
of the Hessian of $\Gamma_{k}\left[\phi\right]+\Delta S_{k}\left[\phi\right]\equiv\tilde{\Gamma}_{k}\left[\phi\right]$:
\begin{eqnarray}
\langle\phi\left(x\right)\phi\left(y\right)\rangle & = & \langle x|\frac{1}{\Gamma_{k}^{\left(2\right)}\left[\phi\right]+R_{k}}|y\rangle\,.\label{eq:C1-martin-corrections}
\end{eqnarray}
We assume that we solved the flow equation and found some RG trajectory
along which we follow the evolution of the two-point function. For
simplicity's sake we focus on the classical regime of the RG trajectory
where we can approximate $G_{k}=\mbox{const}\equiv G$, and in addition
we suppose that the cosmological constant can be neglected, $\Lambda_{k}=0$.
Choosing the flat reference metric $\hat{g}_{\mu\nu}=\delta_{\mu\nu}$
we obtain then, in $d=4$,
\begin{eqnarray}
\langle\phi\left(x\right)\phi\left(y\right)\rangle & = & \langle x|\frac{1}{\left(-\frac{3}{4\pi G}\right)\left(-\Box+{\cal R}_{k}\left(-\Box\right)\right)}|y\rangle\nonumber \\
 & = & \int\frac{d^{4}p}{\left(2\pi\right)^{4}}\frac{1}{\left(-\frac{3}{4\pi G}\right)\left(p^{2}+{\cal R}_{k}\left(p^2 \right)\right)}e^{ip\left(x-y\right)}\,.\label{eq:C2-martin-corrections}
\end{eqnarray}
Clearly, if we set $x=y$ the above integral diverges and the limit
$ \lim_{x\rightarrow y}\,\langle\phi\left(x\right)\phi\left(y\right)\rangle $
is undefined. 
In order to arrive at an expression with more
regular properties we consider the RG running of the two-point function
and take the limit $x\rightarrow y$ only at the level of its scale
derivative, which turns out well defined. Differentiating (\ref{eq:C2-martin-corrections})
we see that the running of the two point function is given by: 
\begin{eqnarray}
\partial_{t}\langle\phi\left(x\right)\phi\left(y\right)\rangle & = & \partial_{t}\langle x|\frac{1}{\tilde{\Gamma}_{k}^{\left(2\right)}}|y\rangle\nonumber \\
 & = & -\langle x|\frac{1}{\tilde{\Gamma}_{k}^{\left(2\right)}}\left(\partial_{t}\tilde{\Gamma}_{k}^{\left(2\right)}\right)\frac{1}{\tilde{\Gamma}_{k}^{\left(2\right)}}|y\rangle=-\langle x|\frac{1}{\tilde{\Gamma}_{k}^{\left(2\right)}}\left(\partial_{t}R_{k}\right)\frac{1}{\tilde{\Gamma}_{k}^{\left(2\right)}}|y\rangle\,.\label{eq:running_2-point_function_1-loop}
\end{eqnarray}
Thus, with our approximations $\Lambda_{k}\approx0$ and $\partial_{t}G_{k}\approx0$,
one obtains:
\begin{eqnarray}
\partial_{t}\langle\phi\left(x\right)\phi\left(y\right)\rangle & = & \frac{4\pi G}{3}\int\frac{d^{4}p}{\left(2\pi\right)^{4}}\frac{e^{ip\left(x-y\right)}}{\left(p^{2}+{\cal R}_{k}\left(p^{2}\right)\right)^{2}}\partial_{t}{\cal R}_{k}\left(p^{2}\right)\,,\label{eq:C3-martin-corrections}
\end{eqnarray}
We immediately notice that the function (\ref{eq:C3-martin-corrections})
is well defined in the limit $x\rightarrow y$ thanks to the presence
of the $k$-derivative of the cutoff kernel ${\cal R}_{k}$. 
For example, employing the optimized cutoff \cite{Litim:2001up} one finds explicitly:
\begin{eqnarray}
\partial_{t}\langle\phi\left(x\right)\phi\left(y\right)\rangle & = & \frac{8\pi G}{3}k^{2}F\left(k\left|x-y\right|\right)\,,\label{eq:C4-martin-corrections}
\end{eqnarray}
with the function $F$ defined by
\begin{eqnarray}
F\left(k\left|x-y\right|\right) & \equiv & \int\frac{d^{4}q}{\left(2\pi\right)^{4}}e^{iq_{\mu}k\left(x-y\right)^{\mu}}\theta\left(1-q^{2}\right)\,.\label{eq:C5-martin-corrections}
\end{eqnarray}

In principle we can now solve for the evolution equation (\ref{eq:C4-martin-corrections})
and obtain the $k$-dependence of the two-point function at arbitrary
points $x$ and $y$. 

In order to find the composite operator of interest
we set $y=x$ in (\ref{eq:C4-martin-corrections}) and obtain
\begin{eqnarray*}
\partial_{t}\langle\phi\left(x\right)^{2}\rangle & = & \frac{8\pi G}{3}k^{2}F\left(0\right)=\frac{1}{12\pi}Gk^{2}\,. \label{eq:20-prime-martin-corrections}
\end{eqnarray*}
Integration leads to the following running correlation function of
the composite operator $\phi^{2}$:
\begin{eqnarray*}
\langle\phi\left(x\right)^{2}\rangle_{k}-\langle\phi\left(x\right)^{2}\rangle_{0} & = & \frac{1}{24\pi}Gk^{2}\,. \label{eq:20-2prime-martin-corrections}
\end{eqnarray*}
Recalling $g_{\mu\nu}=\phi^{2}\hat{g}_{\mu\nu}$, and denoting $\langle\phi\left(x\right)^{2}\rangle_{0}\equiv\tau$
we have the final result
\begin{equation}
 \fbox{$\displaystyle
\langle g_{\mu\nu}\rangle_{k} = \left(1+\frac{1}{\tau}\frac{1}{24\pi}Gk^{2}\right)\langle g_{\mu\nu}\rangle_{0}\,. 
\label{eq:20-3prime-martin-corrections}
$}
\end{equation}

This simple example makes it quite obvious that in general the exploration
of the predictions from the same theory requires much more than merely the scale
dependence of the couplings in the (truncated) EAA, the reason being that there are physically
relevant operators which are not elements of the theory space the EAA lives in, either
as a consequence of a truncation, or even at the exact level.
As we shall see, this complication is particularly acute in quantum gravity because of the complicated 
nature of the observables.

The reader may wonder why we considered the equation for the running
two-point function instead of using directly the ``master equation'' (\ref{eq:flow_eq_composite_operator_epsilon}).
Indeed, as we shall see in a moment, the same results can be obtained
using equation (\ref{eq:flow_eq_composite_operator_epsilon}). 
Employing the two-point function is an instructive alternative though.
It may turn out to be cumbersome however when considering different operators
like $\phi^{4}$, that would require to consider the coincident limit
of a four-point function. 

Now let us turn to the master equation (\ref{eq:flow_eq_composite_operator_epsilon})
and find the running of the composite operator $\phi^{2}$. A simple
one-loop computation yields:
\begin{eqnarray}
\partial_{t}\left[\phi^{2}\left(x\right)\right] & = & -\int\frac{d^{4}p}{\left(2\pi\right)^{4}}\frac{1}{\Gamma_{k}^{\left(2\right)}+{\cal R}_{k}}\frac{1}{\Gamma_{k}^{\left(2\right)}+{\cal R}_{k}}\partial_{t}{\cal R}_{k} \Bigr|_{\hat{g}_{\mu\nu}=\delta_{\mu\nu}, -\Box\rightarrow p^2} \,. 
\label{eq:phi-square-CREH-metric-section}
\end{eqnarray}
Here the factor $1/2$ in the RHS of (\ref{eq:flow_eq_composite_operator_epsilon})
got cancelled by the factor $2$ coming from the Hessian of $\phi^{2}$.
We observe that equation (\ref{eq:phi-square-CREH-metric-section})
is equivalent to equation (\ref{eq:C3-martin-corrections})
in the limit $y=x$ once the truncation (\ref{eq:CREH_truncation_generic_dimension}) is used. 
This equivalence, however, is no longer there
if one goes beyond the one-loop approximation, simply because these two procedures define
different schemes according to which one can renormalize $\phi^{2}$.

Summarizing, this computation shows how one can properly define a
composite metric in the FRG framework, using the CREH truncation as
an example. A similar reasoning will be applied in the following sections
to other composite operators.

\section{Geometric observables: volume and length operators \label{sec:Volume-operators-in-AS}}

In this section we consider the scaling behaviour of two geometrical objects:
(1) the volume operator $ V \equiv \int d^d x\, \sqrt{g} $,
a quantity that one is naturally led to consider as a first possible
observable in quantum gravity, and (2) the length of an arbitrary curve. 
Interestingly, the scaling properties of geometric observables, like the volume and the length,
plays a central role in the description of $2D$ quantum gravity and
have been widely explored \cite{KPZ88,DDK88}.
In this section we consider these geometrical objects in the AS scenario in dimension $d>2$.
We postpone the two dimensional case to section \ref{sec:Two-dimensional-QG}. 

In sections \ref{sub:Volume-operators-in-CREH} and \ref{sub:Volume-operators-in-the-EH-truncation}
we study the volume operator in the CREH truncation, and in the
full-fledged Einstein-Hilbert truncation, respectively. Then, in
section \ref{sub:length-curve-4D}, we investigate the length of a given
spacetime curve in the Einstein-Hilbert truncation.

\subsection{Volume operator in the CREH approximation \label{sub:Volume-operators-in-CREH}}

As we have already seen, in the conformally reduced setting the metric
is a composite operator. Thus any operator $O$ depending on the metric is also a composite operator. 

In $d>2$ dimensions we have the volume element
\begin{eqnarray*}
\sqrt{g} & = & \phi^{d\nu\left(d\right)}\sqrt{\hat{g}}\,.
\end{eqnarray*}
The exponent $d\nu\left(d\right)$ is non-integer except in the dimensions reported in table \ref{table:Conf_Factors_Vol_Operators}.
In two dimensions the exponential parametrization is the distinguished
one leading to a free kinetic term and thus takes the place of the power type dependence $\propto \phi^{2\nu \left(d\right)}$
\cite{Nink:2015lmq}; we shall consider the relevant composite operator
in section \ref{sec:Two-dimensional-QG}.

{ 
\renewcommand{\arraystretch}{1.3} 
\begin{table} 
\begin{center} 
\begin{tabular}{ | c | c | c | c | c | } 
\hline 
$d$   & $3$ & $4$ & $6$ \\ 
\hline\hline 
Conformal factor &  $\phi^4$  &  $\phi^2$  &  $\phi$ \\  
\hline 
Volume operator &  $\phi^6$  &  $\phi^4$  &  $\phi^3$ \\ 
\hline 
\end{tabular} 
\end{center} 
\caption{The composite conformal factors and volume operators for the distinguished parametrizations in various dimensions.} \label{table:Conf_Factors_Vol_Operators} 
\end{table} 
}

We have evaluated the anomalous dimensions of the
volume operators with integer exponents listed in Table \ref{table:Conf_Factors_Vol_Operators}
via equation (\ref{eq:flow_eq_composite_operator_epsilon}), i.e.~those for $d=3,4$ and $6$. 
The calculation makes essential use of equation (\ref{eq:flow_eq_gamma_Z_no_mixing}) and 
it parallels those described in the previous sections, so that it suffices to comment the results. 

First let us consider the case $d=4$, for which $\sqrt{g}=\phi^{4}\sqrt{\hat{g}}$.
The anomalous dimension of the volume operator can be computed expanding
the RHS of equation (\ref{eq:flow_eq_gamma_Z_no_mixing}) up to $\phi^{4}$.
It is easy to observe that the flow equation induces mixing with infinitely
many other operators. We consider a simple non-mixing ansatz, namely
$\left[\phi^{4}\right]=Z_{\phi^{4}}\phi^{4}$, and compute the anomalous
dimension following the discussion of section \ref{sub:Composite-operators-in-FRG}.
The Hessian of the action (\ref{eq:CREH_truncation_generic_dimension})
in four dimensions reads
\begin{eqnarray*}
\Gamma_{k}^{\left(2\right)} & = & \left(-\frac{3}{4\pi G_{k}}\right)\left(-\hat{\Box}+\frac{\hat{R}}{6}-2\Lambda_{k}\phi^{2}\right)\,.
\end{eqnarray*}
Inserting the Hessian $\Gamma_{k}^{\left(2\right)}$ and the Hessian
of the operators $\left[\phi^{4}\right]$ in equation (\ref{eq:flow_eq_gamma_Z_no_mixing})
one can read off the anomalous dimension. The correction to the classical
scaling dimension associated to the volume operator can be found in
table \ref{table:Vol_Operators_anomalous_dim} together with the cases
for $d=3$ and $d=6$.

{ 
\renewcommand{\arraystretch}{1.3} 
\begin{table} 
\begin{center} 
\begin{tabular}{ | c | c | c | c | c | } 
\hline 
$d$   & $3$ & $4$ & $6$ \\ 
\hline\hline 
Volume operator &  $\phi^6$  &  $\phi^4$  &  $\phi^3$ \\ 
\hline 
anomalous dimension &  $ 0 $  &  $\frac{2}{\pi} G \Lambda$  &  $\frac{27}{250 \pi ^2} G \Lambda ^2$ \\ 
\hline 
\end{tabular} 
\end{center} 
\caption{Volume operators in various dimensions and their one-loop anomalous dimensions according to the CREH model.} \label{table:Vol_Operators_anomalous_dim} 
\end{table} 
}

From table \ref{table:Vol_Operators_anomalous_dim} we note that the
anomalous dimensions are proportional to the Newton's constant and
a certain power of the cosmological constant that renders $\gamma_V$ dimensionless. 
The factor of $\Lambda$ comes from expanding, in the field $\phi$, the regularized propagator
in the flow equation; the power of $\Lambda$ is essentially determined
by the order in this expansion. The three dimensional case shows a
vanishing anomalous dimension in our approximation.\footnote{
In the three dimensions the anomalous dimension vanishes if
we parametrize $\left[\phi^{6}\right]=Z_{6}\phi^{6}$. A non-zero
anomalous dimension occurs upon including the mixing of $\phi^6$ with other operators.}

\subsection{Volume in the full Einstein-Hilbert truncation \label{sub:Volume-operators-in-the-EH-truncation}}

In this section we estimate the anomalous dimension of the volume
operator via a non-mixing ansatz within the fully fledged Einstein-Hilbert
truncation for the gravitational EAA:
\begin{eqnarray}
\Gamma_k \left[g_{\mu\nu}\right] & = & \frac{1}{16\pi G_k}\int d^{d}x\,\sqrt{g}\left(2\Lambda_k -R\right) \,.\label{eq:EH_truncation_full}
\end{eqnarray}
The metric $g_{\mu\nu}$ is expressed via the sum of a background metric $\bar{g}_{\mu\nu}$ and
the dynamical metric $h_{\mu\nu}$, i.e.~$g_{\mu\nu}=\bar{g}_{\mu\nu}+h_{\mu\nu}$.
We equip the ansatz (\ref{eq:EH_truncation_full}) with the Feynman-de
Donder gauge fixing, which gives a particularly simple Hessian, see
for instance \cite{CPR09}. 

We consider now the integrated volume operator $V=\int d^{d}x\sqrt{g}$, and we
do not allow for any mixing with other operators. 
In order to compute the associated anomalous dimension,
and thus the scaling properties of $V$, we employ equation (\ref{eq:flow_eq_gamma_Z_no_mixing}).

Taking into account the presence of the ghosts, the Hessian of the
functional $V\equiv V\left[g\right]\equiv V\left[\bar{g}+h\right]$ has the following block form in field space:
\begin{eqnarray*}
V^{\left(2\right)} & = & \left(\begin{array}{ccc}
\frac{\delta^{2} V}{\delta h\delta h} & 0 & 0\\
0 & 0 & 0\\
0 & 0 & 0
\end{array}\right)\,.
\end{eqnarray*}
Due to the simple structure of $V^{\left(2\right)}$, 
equation (\ref{eq:flow_eq_gamma_Z_no_mixing}) can be expressed
solely in terms of the regularized graviton propagator, i.e.
\begin{eqnarray}
\gamma_V V & = & -\frac{1}{2}\mbox{Tr}\left[\frac{1}{\Gamma_{k,hh}^{\left(2\right)}+{\cal R}_{k,hh}}\cdot\left(\frac{\delta^{2} V}{\delta h\delta h}\right)\cdot\frac{1}{\Gamma_{k,hh}^{\left(2\right)}+{\cal R}_{k,hh}}\cdot\partial_{t} {\cal R}_{k}\right]\,,\label{eq:flow_eq_gamma_vol_operator_full_EH}
\end{eqnarray}
where all quantities in (\ref{eq:flow_eq_gamma_vol_operator_full_EH}) are evaluated at $h=0$ now. Hence in particular
\begin{eqnarray}
\left( \frac{\delta^{2} V}{\delta h\delta h} \right)_{\mu\nu}^{\quad \rho\sigma} 
& = & \left(-\frac{1}{2}\sqrt{g}\right)\left(\mathbb{I}_{\mu\nu}^{\,\,\,\,\,\,\,\rho\sigma}-\frac{d}{2}\mathbb{P}_{\mu\nu}^{\,\,\,\,\,\,\,\rho\sigma}\right)\,,\label{eq:Hessian_vol_operator_full}
\end{eqnarray}
with the matrices
$
\mathbb{I}_{\mu\nu}^{\,\,\,\,\,\,\,\rho\sigma}  \equiv  \frac{1}{2}\left(\delta_{\mu}^{\rho}\delta_{\nu}^{\sigma}+\delta_{\mu}^{\sigma}\delta_{\nu}^{\rho}\right)
$
and
$
\mathbb{P}_{\mu\nu}^{\,\,\,\,\,\,\,\rho\sigma}  \equiv  \frac{1}{d}g_{\mu\nu}g^{\rho\sigma}\,.
$
Inserting the operator (\ref{eq:Hessian_vol_operator_full}) in equation
(\ref{eq:flow_eq_gamma_vol_operator_full_EH}) one finds
\begin{eqnarray}
\gamma_{V} & = & \frac{d\left(d+1\right)}{2}\frac{16\pi G_{k}}{2}
\left[\frac{1}{\left(4\pi\right)^{d/2}}\frac{1}{\Gamma\left(\frac{d}{2}\right)}\int_{0}^{\infty}dz\,\frac{z^{d/2-1}}{\left(z+R_{k}\right)^{2}}\left(\partial_{t}R_{k}-\frac{\partial_{t}G_{k}}{G_{k}}R_{k}\right)\right]\,.\label{eq:anomalous_dim_volume_EH_truncation}
\end{eqnarray}
The integral can be evaluated in terms of the standard threshold functions
$\Phi_{n}^{p}$ and $\tilde{\Phi}_{n}^{p}$ from \cite{R98}. In terms
of the dimensionless couplings $g_{k}\equiv k^{d-2}G_{k}$ and $\lambda_{k}\equiv\Lambda_{k}/k^{2}$,
and with the anomalous dimension related to the Newton's constant,
$\eta_{N}\equiv\partial_{t}G_{k}/G_{k}$, we find 
\begin{equation}
 \fbox{$\displaystyle
\gamma_{V}\left(g,\lambda\right) =  d\left(d+1\right)\frac{g}{\left(4\pi\right)^{d/2-1}}\left[\Phi_{d/2}^{2}\left(-2\lambda\right)-\eta_{N}\left(g,\lambda\right)\tilde{\Phi}_{d/2}^{2}\left(-2\lambda\right)\right]\,. \label{eq:18-1-martin-corrections}
  $}
 \end{equation}
This formula applies to an arbitrary point of the ($g$-$\lambda$)--theory
space. For $\eta_{N}\left(g,\lambda\right)$ one should substitute
the standard result from the (full) Einstein-Hilbert truncation.\footnote{See eqs.~(4.41) with (4.40) in \cite{R98}.}

For the example of the optimized cutoff, equation (\ref{eq:18-1-martin-corrections})
becomes:
\begin{eqnarray}
\gamma_{V}\left(g,\lambda\right) & = & \frac{4\left(d+1\right)}{\left(4\pi\right)^{d/2-1}\Gamma\left(\frac{d}{2}\right)}\frac{g}{\left(1-2\lambda\right)^{2}}\left(1-\eta_{N}\left(g,\lambda\right)\frac{1}{d+2}\right)\,.\label{eq:18-2-martin-corrections}
\end{eqnarray}
At the NGFP $\left(g_*,\lambda_*\right)$ where $\eta_{N}\left(g_*,\lambda_*\right)=2-d$, equation
(\ref{eq:18-2-martin-corrections}) yields for $\gamma_{V}^{*}=\gamma_{V}\left(g_{*},\lambda_{*}\right)$:
\begin{eqnarray}
\gamma_{V}^{*} & = & \frac{8d\left(d+1\right)}{\left(4\pi\right)^{d/2-1}\Gamma\left(\frac{d}{2}\right)\left(d+2\right)}\frac{g_{*}}{\left(1-2\lambda_{*}\right)^{2}}\,.\label{eq:18-3-martin-corrections}
\end{eqnarray}

For a first orientation, let us focus on $d=2+\varepsilon$ dimensions where
the Einstein-Hilbert truncation is known \cite{R98} to display a non-Gaussian
fixed point which has the (universal) coordinate $g_{*}=\frac{3}{38}\varepsilon$,
together with a non-universal $\lambda_{*}$ which is likewise of order
$\varepsilon$. Since $\Phi_{1}^{2}\left(0\right)=1$ for any cutoff
shape function, equation (\ref{eq:18-1-martin-corrections}) yields
in this case
\begin{eqnarray}
\gamma_{V}^{*} & = & 12g_{*}+O\left(\varepsilon^{2}\right)=\frac{18}{19}\varepsilon+ O\left(\varepsilon^{2}\right)\,.\label{eq:18-4-martin-corrections}
\end{eqnarray}
This anomalous dimension amounts to a shift of the classical scaling dimension of the volume
operator, $d_{V}=-d$, to the corrected value $d_{V}+\gamma_{V}^{*}=-2-\varepsilon+\frac{18}{19}\varepsilon=-2-\frac{1}{19}\varepsilon$,
which appears to correspond to an effective spacetime dimensionality
which is slightly smaller (larger) than the classical one when $\varepsilon>0$
($\varepsilon<0$).

The value of $\gamma_{V}$ at the UV fixed point for a four dimensional
spacetime is reported in Table \ref{table:length_and_Vol_anomalous_dim}
in section \ref{sub:length-curve-4D}. 
It is worth to notice that the value of this anomalous dimension, $\gamma_V^* \approx 3.9866$, is almost equal to the spacetime
dimension, i.e.~$\gamma_{V}^*\approx4$, and that the volume operator
has classical {\it mass} dimension $d_{V}=-4$. 
According to the discussion and conventions
of section \ref{sub:Composite-operators-in-FRG}, the full scaling
dimension of the volume operator (and analogously for any other operator)
is obtained adding the anomalous dimension to the classical mass dimension: $d_V^{\rm{corrected}}=d_V+\gamma_V^*$. 
In the present case the
quantum contribution almost cancels against the classical value 
so that the operator $V$ has
{\it an almost vanishing scaling dimension}, $d_V+\gamma_V^* \approx 0$.
 
Let us stress that this result may well be an artifact of the truncation
and approximations employed so far. Nevertheless, we could possibly make
contact with independent results in the literature. 
First we recall the
connection between the ``composite operator formalism'' of section
\ref{sub:Composite-operators-in-FRG} with that of critical exponents
$\theta_i$ defined by the linearized flow.
In particular, equation (\ref{eq:connection_critical_exp_anomalous_dim})
allows one to compare our results with those obtained by linearizing
the RG around the fixed point. 
Unfortunately, most of the works in the Asymptotic
Safety literature have produced complex critical exponents $\theta$
so far and thus a direct comparison with our present results is far from obvious.

Therefore, at least within our approximations,
there is an indication which hints towards {\it a perfect cancellation between
classical and quantum contributions to the scaling dimension of the
volume operator}. 
If it were so, this could suggest that at very small distance
scales (fixed point regime) the spacetime is actually much more empty
than one would naively expect.

\subsection{Quantum average of the length of curves \label{sub:length-curve-4D}}

In this section we study another geometrical object: the length of
curves. Let us denote $x^{\mu}\left(s\right)$ the coordinates of the points visited by
a curve as a function the parameter $s\in\left[0,1\right]$. The length
of this curve on a manifold of fixed metric $g_{\mu\nu}$ is then given by
\begin{eqnarray}
L\left[x\left(\cdot\right),g\right] & \equiv & \int_{0}^{1}ds\sqrt{g_{\mu\nu}\left(x\left(s\right)\right)\dot{x}^{\mu}\left(s\right)\dot{x}^{\nu}\left(s\right)}\,,\,\dot{x}^{\mu}\left(s\right)\equiv\frac{dx^{\mu}\left(s\right)}{ds}\,.\label{eq:27-prime-martin-corrections}
\end{eqnarray}
One is interested in quantum averages of $L\left[x\left(\cdot\right),g\right]$
over the metrics $g$ realized on the manifold,
\begin{equation}
\langle\int_{0}^{1}ds\sqrt{g_{\mu\nu}\left(x\left(s\right)\right)\dot{x}^{\mu}\left(s\right)\dot{x}^{\nu}\left(s\right)}\,\rangle\,.
\end{equation}
With regard to the quantum metric, $g_{\mu\nu}$, or rather the fluctuation $h_{\mu\nu}=g_{\mu\nu}-\bar{g}_{\mu\nu}$, 
which is considered an elementary field here,
the length $L \left[x\left(\cdot \right), g \right] \equiv L \left[x\left(\cdot \right), \bar{g}+h \right] $
is clearly a composite operator. 
It is therefore natural to ask if
this operator possesses a non-trivial anomalous dimension which encodes
how the length responds to a scale variation. 

As usual, we shall compute the anomalous dimension via the ``master equation'' (\ref{eq:flow_eq_composite_operator_epsilon}).
We need the Hessian of the functional $L$ with respect to
the metric, for a fixed curve $x^{\mu}\left(s\right)$,
on the RHS of equation (\ref{eq:flow_eq_composite_operator_epsilon}).
It reads explicitly
\begin{equation}
\frac{\delta^{2}L\left[x\left(\cdot\right),g\right]}{\delta h_{\alpha\beta}\left(x^{\prime}\right)\delta h_{\gamma\delta}\left(x^{\prime\prime}\right)}  = 
 -\frac{1}{4}\int ds\,\frac{\dot{x}^{\alpha}\left(s\right)\dot{x}^{\beta}\left(s\right)\dot{x}^{\gamma}\left(s\right)\dot{x}^{\delta}\left(s\right)}{\left[g_{\mu\nu}\left(x_{\lambda}\right)\dot{x}^{\mu}\left(s\right)\dot{x}^{\nu}\left(s\right)\right]^{3/2}}\delta\left(x^{\prime}-x\left(s\right)\right)\delta\left(x^{\prime\prime}-x\left(s\right)\right).  \quad \quad
\label{eq:27-prime-red-martin-corrections}
\end{equation}
We can anticipate that by taking the trace in the flow equation we
are lead to contract the indices of the Hessian (\ref{eq:27-prime-red-martin-corrections})
in such a way that the RHS turns out to be proportional to $L$ itself.
Considering a flat spacetime background metric
it is rather straightforward to obtain the associated anomalous
dimension $\gamma_{L}$ from equation (\ref{eq:flow_eq_gamma_Z_no_mixing}).

The final result obtained in this way reads
\begin{equation}
 \fbox{$\displaystyle
\gamma_{L}\left(g,\lambda\right)  =  \frac{d-3}{d-2}\frac{g}{\left(4\pi\right)^{d/2-1}}\left[\Phi_{d/2}^{2}\left(-2\lambda\right)-\eta_{N}\left(g,\lambda\right)\tilde{\Phi}_{d/2}^{2}\left(-2\lambda\right)\right]\,.\label{eq:L-1-martin-corrections}
  $}
 \end{equation}
The corresponding one-loop result could be retrieved by neglecting
the term proportional to $\eta_{N}\left(g,\lambda\right)$ on the
RHS of (\ref{eq:L-1-martin-corrections}) and letting $\lambda\rightarrow0$
in the argument of the threshold functions. 

Interestingly enough, the function $\gamma_{L}\left(g,\lambda\right)$
is proportional to $\gamma_{V}\left(g,\lambda\right)$. At any point
$\left(g,\lambda\right)$, and for all cutoff functions we have
\begin{eqnarray}
\gamma_{L}\left(g,\lambda\right) & = & \frac{d-3}{d\left(d+1\right)\left(d-2\right)}
\, \gamma_{V} \left(g,\lambda\right)\,.\label{eq:L-2-martin-corrections}
\end{eqnarray}
Thus, for example, $\gamma_{L}=\gamma_{V}$ in $d=1$, and $\gamma_{L}=\frac{1}{40}\gamma_{V}$
in $d=4$, everywhere in the theory space. 

As for the $(2+\varepsilon)$-dimensional case, it is remarkable that
the pole proportional to $1/\left(d-2\right)$ in (\ref{eq:L-2-martin-corrections})
cancels the linear $\varepsilon$-dependence of $\gamma_{V}^{*}$.
Hence, with (\ref{eq:18-4-martin-corrections}) in the leading order
in $\varepsilon$,
\begin{eqnarray*}
\gamma_{L}^{*} & = & -\frac{1}{6\varepsilon}\gamma_{V}^{*}=-\frac{2 g_{*}}{\varepsilon}\,.
\end{eqnarray*}
This yields a finite, non-zero anomalous dimension in the limit $\varepsilon\rightarrow0$:
\begin{eqnarray*}
\gamma_{L}^{*} & = & -\frac{3}{19}\,.
\end{eqnarray*}
Recall that the anomalous dimension of the volume operator vanishes
in this limit, $\gamma_{V}^{*}=0+O\left(\varepsilon\right)$. 

The numerical results for the four dimensional case can be found in Table \ref{table:length_and_Vol_anomalous_dim}.
They were obtained with the optimized cutoff \cite{Litim:2001up}.
Note that the full anomalous dimensions $\gamma_L^*$ and $\gamma_V^*$ do indeed differ
by the universal factor $1/40$ predicted above.

{ 
\renewcommand{\arraystretch}{1.3} 
\begin{table} 
\begin{center} 
\begin{tabular}{ | c | c | c | } 
\hline 
     & $\gamma_{L}^{*}$ & $\gamma_{V}^{*}$ \\ 
\hline\hline 
one-loop anomalous dimension &  $0.0682$  &  $2.7273$   \\ 
\hline 
full anomalous dimension     &  $ 0.0997 $     &  $3.9866$   \\ 
\hline 
\end{tabular} 
\end{center} 
\caption{The anomalous dimensions of the length and volume operators in $d=4$.} \label{table:length_and_Vol_anomalous_dim} 
\end{table} 
}

\section{Quantum gravity in exactly two dimensions \label{sec:Two-dimensional-QG}}

Liouville field theory is a well known playground for quantum gravity
in (exactly) two dimension \cite{KPZ88,DDK88}. 
Along a different line
of investigations, Quantum Einstein Gravity (QEG) in $2+\varepsilon$
dimensions has often been used as a theoretical laboratory for Asymptotic
Safety. There, $\varepsilon$ is always kept different from zero
since, if one employs the Einstein-Hilbert truncation, the (bare)
action becomes purely topological at $\varepsilon=0$. 

However, recently it has been shown that if one takes the limit $\varepsilon\rightarrow0$ of
the action functional only {\it after} having already computed the
RG flow in $2+\varepsilon$ dimensions, one obtains a non-trivial
EAA and fixed point action \cite{Nink:2015lmq}. 
The latter action is given by the following manifestly $2$-dimensional functional 
which has the form of
the induced gravity action: 
\begin{eqnarray}
\Gamma_{k\rightarrow\infty} & = & -\frac{\left(25-N\right)}{96\pi}\int d^{2}x\sqrt{g}R \left(-\frac{1}{\Box} \right) R+\cdots\,.\label{eq:2D-FP-EAA}
\end{eqnarray}
Here the dots stand for a cosmological constant term with a non-universal
coefficient. This result applies to gravity coupled to $N$ minimally
coupled free scalars fields, and the exponential parametrization of
the metric fluctuations; for the standard linear parametrization,
the central charge $25$ in equation (\ref{eq:2D-FP-EAA}) would be replaced
by $19$, see \cite{Nink:2014yya}.

The $2D$ functional (\ref{eq:2D-FP-EAA}) descends from the $\left(2+\varepsilon \right)$-dimensional
Einstein-Hilbert term alone.
Therefore the total EAA contains further contributions, in particular the Faddeev-Popov ghosts and the Jacobian
leading to a Weyl invariant measure.
These contributions change the functional in (\ref{eq:2D-FP-EAA}), yielding an exactly vanishing total charge of
QEG in $2D$. 
(For further details we refer to \cite{Nink:2015lmq}.)
In the following we shall consider the contribution (\ref{eq:2D-FP-EAA}) in its own right though.

Inserting metrics of the form $g_{\mu\nu}=e^{2\phi}\hat{g}_{\mu\nu}$,
the action (\ref{eq:2D-FP-EAA}) gives rise to a Liouville theory
for $\phi$:
\begin{eqnarray}
\Gamma_{k\rightarrow\infty} & = & -\frac{\left(25-N\right)}{24\pi}\int d^{2}x\sqrt{\hat{g}}\left\{ \phi 
\left(-\hat{\Box} \right) \phi+\hat{R}\phi+\mu_{*}k^{2}e^{2\phi}\right\} \,.\label{eq:30-prime-martin-corrections}
\end{eqnarray}
It describes a RG fixed point\footnote{In the notation of \cite{Nink:2015lmq}, $\mu_{*}\equiv-2\breve{\lambda}_{*}\equiv-2\lambda_{*}/\varepsilon$.}
on the side of the {\it effective} action, in particular $\phi\equiv\langle\chi\rangle$
is the expectation value of the quantum field, $\chi$. Furthermore,
in \cite{Nink:2015lmq,NR16} also the (re)construction of a well defined
UV-regularized functional integral $\int{\cal D}_{\Lambda}\chi\,e^{-S_{\Lambda}\left[\chi\right]}$
has been performed which reproduces the RG trajectories $\Gamma_{k}\left[\phi\right]$.

Employing the approach outlined in \cite{Manrique:2008zw}, and the UV-regularized
measure proposed there, the dependence of the bare action $S_{\Lambda}\left[\chi\right]$
on the UV cutoff scale $\Lambda$ was deduced from (\ref{eq:30-prime-martin-corrections}),
with the result 
\begin{eqnarray}
S_{\Lambda\rightarrow\infty} & = & \kappa\int d^{2}x\sqrt{\hat{g}}\left[\chi\left(-\hat{\Box}\right)\chi+\hat{R}\chi+\check{\mu}_{\Lambda}\Lambda^{2}e^{2\chi}\right]\,.\label{eq:2D-FP-action}
\end{eqnarray}
Remarkably, the coefficient of the bare kinetic term turned out to
be exactly the same as its counterpart at the effective level, namely
\begin{equation}
\kappa=-\frac{\left(25-N\right)}{24\pi}\,.\label{eq:30-third-martin-corrections}
\end{equation}
The bare fixed point cosmological constant $\check{\mu}_{*}$ is different
from the effective one, $\mu_{*}$, and depends on the precise definition
of the measure, ${\cal D}_{\Lambda}\chi$. There exists a normalization
such that the bare cosmological constant vanishes. We take advantage
of this possibility and henceforth set $\check{\mu}_{*}=0$. For the
details of the reconstruction step we must refer to the literature
\cite{Manrique:2008zw,Nink:2015lmq,NR16}.

In this context, Liouville theory comes into play as the exactly two
dimensional limiting case of $d$-dimensional QEG, as always based
on the functional integral $\int{\cal D}_{\Lambda}g_{\mu\nu}^{\rm bare}\,e^{-S_{\Lambda}\left[g_{\mu\nu}^{\rm bare}\right]}$,
but now only over metrics of the type 
\begin{equation}
g_{\mu\nu}^{\rm bare}\equiv e^{2\chi}\hat{g}_{\mu\nu}\,.\label{eq:31-martin-corrections}
\end{equation}

In the present paper instead we shall not be concerned with the physical origin of
the Liouville theory, and rather use it as a framework to see
the FRGE for composite operators ``at work'' and to show how it
relates to the standard approaches. We shall employ the action (\ref{eq:2D-FP-action})
with an arbitrary value of $\kappa$ and, for simplicity, $\check{\mu}_{*}=0$.

\subsection{Correlators of exponential operators}

From the splitting (\ref{eq:31-martin-corrections}) it is clear that
with respect to the elementary field, the bare conformal factor $\chi$, the metric 
is a composite operator. 
We are interested in evaluating its correlation functions
which boils down to evaluating correlators of ``vertex operators'': 
\begin{equation}
\langle e^{2a_{1}\chi\left(x_{1}\right)}\cdots e^{2a_{n}\chi\left(x_{n}\right)}\rangle\,.\label{eq:correlation_of_multiple_exp2a_iChi}
\end{equation}
Let us consider a flat background metric, implying $\hat{R}=0$, and
take $\mu_{\Lambda}=0$. \\
$\boldsymbol{\left(1\right)}$
First we focus on the one point function $\langle e^{2a_{1}\chi\left(x\right)}\rangle$.
The scale dependence of this average could be straightforwardly found
employing the techniques used in sections \ref{sec:Composite-metrics-in-the-CREH}
and \ref{sec:Volume-operators-in-AS}. However, here
it turns out more convenient to work directly at the path integral
level rather than working out the ``master equation'' at higher order in
the source $\varepsilon$. Thus we first consider 
\begin{eqnarray*}
\langle e^{2a\chi\left(x\right)}\rangle_{k} & = & \frac{1}{Z_{0}}\int{\cal D}\chi\,\exp\left\{ -\kappa\int d^2y\,\chi\left(-\Box+{\cal R}_{k}\right)\chi+2a\chi\left(x\right)\right\} \\
 & \equiv & \frac{1}{Z_{0}}\int{\cal D}\chi\,\exp\left\{ -\kappa\int d^2y\,\chi\left(-\Box+{\cal R}_{k}\right)\chi+\int d^2y\,J\left(y\right)\chi\left(y\right)\right\} \,,
\end{eqnarray*}
with $J\left(y\right)\equiv2a\delta\left(y-x\right)$. This is a simple
Gaussian integral and so one obtains
\begin{eqnarray}
\langle e^{2a\chi\left(x\right)}\rangle & = & \exp\left[\frac{a^{2}}{\kappa}\langle x|\frac{1}{-\Box+{\cal R}_{k}}|x\rangle\right]=\exp\left[\frac{a^{2}}{\kappa}{\cal G}_{k}\left(0\right)\right]\,,\label{eq:Exp_2aChi_via_G0}
\end{eqnarray}
where ${\cal G}_{k}\left(0\right)=\langle x|\left(-\Box+{\cal R}_{k}\right)^{-1}|x\rangle$
is the Green function at coinciding points. 

Clearly ${\cal G}_{k}\left(0\right)$ is undefined as it stands
and we need a regularization procedure. Rather than the Green function
per see, we determine its scale derivative:
\begin{eqnarray}
\partial_{t}{\cal G}_{k}\left(0\right) & = & =-\langle x|\frac{\partial_{t}{\cal R}_{k}}{\left(-\Box+{\cal R}_{k}\right)^{2}}|x\rangle=-\int\frac{d^{2}q}{(2\pi)^2}\frac{\partial_{t}{\cal R}_{k}\left(q^{2}\right)}{\left(q^{2}+{\cal R}_{k}\left(q^{2}\right)\right)^{2}}\\
 & = & -\frac{1}{2\pi}\Phi_{1}^{2}\left(0\right)=-\frac{1}{2\pi}\,. \label{eq:38prime-martin-corrections}
\end{eqnarray}
The above result is {\it universal} in the sense that $\Phi_{1}^{2}\left(0\right)=1$
is known to be valid for any cutoff of the type ${\cal R}_{k}=k^{2}R^{\left(0\right)}\left(-\Box/k^{2}\right)$
\cite{R98}. By integrating (\ref{eq:38prime-martin-corrections}) we find
\begin{eqnarray}
{\cal G}_{k}\left(0\right)-{\cal G}_{\mu}\left(0\right) & = & -\frac{1}{2\pi}\log\frac{k}{\mu}\,.\label{eq:34-second-martin-corrections}
\end{eqnarray}
Using this result in equation (\ref{eq:Exp_2aChi_via_G0}) we can
form the well defined ratio
\begin{eqnarray*}
\frac{\langle e^{2a\chi\left(x\right)}\rangle_{k}}{\langle e^{2a\chi\left(x\right)}\rangle_{\mu}} & = & \left(\frac{k}{\mu}\right)^{-\frac{1}{2\pi}\frac{a^{2}}{\kappa}}
\end{eqnarray*}
and so we obtain
\begin{eqnarray}
\langle e^{2a\chi\left(x\right)}\rangle_{k} & = & \left(\frac{k}{\mu}\right)^{-\frac{1}{2\pi}\frac{a^{2}}{\kappa}}\langle e^{2a\chi\left(x\right)}\rangle_{\mu}\,.\label{eq:34-one-point-martin-corrections}
\end{eqnarray}
From this relation we can read off the scaling dimension of the exponential
operator:
\begin{equation}
 \fbox{$\displaystyle
\partial_{t}\langle e^{2a\chi\left(x\right)}\rangle_{k}  =  -\frac{1}{2\pi}\frac{a^{2}}{\kappa}\langle e^{2a\chi\left(x\right)}\rangle_{k}\,.\label{eq:1-point-exp-operator-RG}
  $}
 \end{equation}
\\
$\boldsymbol{\left(2\right)}$
Let us compare the $k$-dependence of the expectation value (\ref{eq:34-one-point-martin-corrections}) with the one
found directly from the composite operator flow equation. We denote
$Z_{O}$ the renormalization constant associated to the operator $O\left(x\right)=e^{2a\chi\left(x\right)}$.
A one-loop computation based upon equation (\ref{eq:flow_eq_gamma_Z_no_mixing})
yields then:
\begin{eqnarray}
\left(Z_{O}^{-1}\partial_{t}Z_{O}\right)e^{2a\chi\left(x\right)} & = & 
\mbox{Tr}\left[-\frac{1}{2}\frac{1}{2\kappa\left(-\Box+{\cal R}_{k}\right)}\left(4a^{2}e^{2a\chi\left(x\right)}\right)\frac{1}{2\kappa\left(-\Box+{\cal R}_{k}\right)}2\kappa\partial_{t} {\cal R}_{k}\right] \nonumber \\
 & = & -\left(\frac{a^{2}}{\kappa}\right)\frac{1}{2\pi}\Phi_{1}^{2}\left(0\right)\,e^{2a\chi\left(x\right)} \nonumber \\
 & = & -\frac{1}{2\pi}\left(\frac{a^{2}}{\kappa}\right)\,e^{2a\chi\left(x\right)}\,.  \label{eq:running-vertex-via-FRGE-compOp}
\end{eqnarray}
The running found in equation (\ref{eq:running-vertex-via-FRGE-compOp}) is the expected result, 
the same as in equation (\ref{eq:1-point-exp-operator-RG}).

Furthermore, we mention that this approach, based upon the ``master equation'' (\ref{eq:flow_eq_composite_operator_epsilon}), 
allows one to obtain the so
called KPZ scaling relations in the FRG framework.
As a further illustration of our techniques we discuss their derivation in the Appendix.
For a detailed discussion of Liouville theory and the KPZ scaling in the EAA approach see also \cite{Reuter:1996eg,Codello:2014wfa}.
\\
$\boldsymbol{\left(3\right)}$
Now we generalize (\ref{eq:Exp_2aChi_via_G0}) to the $n$-point correlation functions,
starting out from: 
\begin{eqnarray}
\langle e^{2a_{1}\chi\left(x_{1}\right)}\cdots e^{2a_{n}\chi\left(x_{n}\right)}\rangle_{k} & = & \frac{1}{Z_{0}}\int{\cal D}\chi\,\exp\left\{ -\kappa\int dy\,\chi\left(-\Box+{\cal R}_{k}\right)\chi+\int dy\,J\left(y\right)\chi\left(y\right)\right\} , \nonumber
\end{eqnarray}
with the source function
$ J\left(y\right)\equiv\sum_{i=1}^{n}2a_{i}\delta\left(y-x_{i}\right) $.
The Gaussian integral yields
\begin{eqnarray*}
\langle e^{2a_{1}\chi\left(x_{1}\right)}\cdots e^{2a_{n}\chi\left(x_{n}\right)}\rangle_{k} & = & \exp\left[\sum_{i=1}^{n}\frac{a_{i}^{2}}{\kappa}{\cal G}_{k}\left(0\right)+\sum_{i<j}\frac{2a_{i}a_{j}}{\kappa}{\cal G}_{k}\left(x_{i}-x_{j}\right)\right]\,.
\end{eqnarray*}
While these correlation functions are ill defined their flow equation
is perfectly regular:
\begin{eqnarray}
\partial_{t}\log\langle e^{2a_{1}\chi\left(x_{1}\right)}\cdots e^{2a_{n}\chi\left(x_{n}\right)}\rangle_{k}  =  \sum_{i=1}^{n}\frac{a_{i}^{2}}{\kappa}\partial_{t}{\cal G}_{k}\left(0\right)+\sum_{i<j}\frac{2a_{i}a_{j}}{\kappa}\partial_{t}{\cal G}_{k}\left(x_{i}-x_{j}\right)\label{eq:I4-1-martin-corrections}\\
\equiv  \frac{1}{\kappa}\left(\sum_{i=1}^{n}a_{i}\right)^{2}\partial_{t}{\cal G}_{k}\left(0\right)+\frac{2}{\kappa}\sum_{i<j}a_{i}a_{j}\left[\partial_{t}{\cal G}_{k}\left(x_{i}-x_{j}\right)-\partial_{t}{\cal G}_{k}\left(0\right)\right] \,. 
  \nonumber 
\end{eqnarray}
Here the regularized propagator occurs at non-coincident points also: 
\begin{eqnarray*}
{\cal G}_{k}\left(r\right) & = & \int\frac{d^{2}q}{\left(2\pi\right)^{2}}\,\frac{e^{iq\left(x-y\right)}}{q^{2}+{\cal R}_{k}}\,, \, r\equiv\left|x-y\right|\,.
\end{eqnarray*}
Considering the example of a mass like cutoff profile, i.e.~${\cal R}_{k}=k^{2}$,
we find
\begin{equation}
{\cal G}_{k}\left(r\right) = \frac{1}{2\pi}K_{0}\left(kr\right) \,\,\, \mbox{and} \,\,\,
\partial_{t}{\cal G}_{k}\left(r\right) = -\frac{1}{2\pi}krK_{1}\left(kr\right)\,,
\end{equation}
where $K_{\nu}$ denotes the Bessel function of the second kind. In
the limit $kr\rightarrow0$, it gives
\begin{equation}
{\cal G}_{k}\left(r\right) \approx -\frac{1}{2\pi}\log\left(kr\right)  \mbox{  and   }
\partial_{t}{\cal G}_{k}\left(r\right) \approx -\frac{1}{2\pi}=\mbox{const}\,.
\end{equation}
In particular we recover (\ref{eq:34-second-martin-corrections}),
i.e.,
\begin{eqnarray*}
{\cal G}_{k}\left(0\right)-{\cal G}_{\mu}\left(0\right) & \equiv & \lim_{r\rightarrow0}\left({\cal G}_{k}-{\cal G}_{\mu}\right)\left(r\right)=-\frac{1}{2\pi}\log\left(\frac{k}{\mu}\right)\,.
\end{eqnarray*}
We also see that 
\begin{equation}
\partial_{t}{\cal G}_{k}\left(r\right)-\partial_{t}{\cal G}_{\mu}\left(0\right) \rightarrow 0 \mbox{  for  }kr\rightarrow 0
\end{equation}
is a well defined limit. Hence the last term in equation (\ref{eq:I4-1-martin-corrections})
vanishes when all distances $\left|x_{i}-x_{j}\right|$ are much smaller
than $k^{-1}$.

Let us consider the case $n=2$, for example. At small distances $\left|x_{1}-x_{2}\right|\ll k^{-1}$,
the RG equation (\ref{eq:I4-1-martin-corrections}) yields
\begin{eqnarray}
\partial_{t}\log\langle e^{2a_{1}\chi\left(x_{1}\right)}e^{2a_{2}\chi\left(x_{2}\right)}\rangle_{k} & = & -\frac{1}{2\pi\kappa}\left(a_{1}+a_{2}\right)^{2}\,.\label{eq:I5-1-martin-corrections}
\end{eqnarray}
Note that in the limit $k\rightarrow0$ the distances $\left|x_{1}-x_{2}\right|$,
being measured with the fixed metric $\hat{g}_{\mu\nu}=\delta_{\mu\nu}$,
all become small in comparison with $k^{-1}$. Hence the scaling exponent
displayed by the RHS of (\ref{eq:I5-1-martin-corrections}) is indeed
the expected, correct, and universal result \cite{AADZ94}.

According to (\ref{eq:I5-1-martin-corrections}), the two-point correlator
equals $k^{-\left(a_{1}+a_{2}\right)^{2}/2\pi\kappa}$, multiplied
by a $k$-independent function of $\left|x_{1}-x_{2}\right|$.
To find it, we can start from the formal expression 
\begin{eqnarray*}
\langle e^{2a_{1}\chi\left(x_{1}\right)}e^{2a_{2}\chi\left(x_{2}\right)}\rangle_{k} & = & \exp\left[\frac{a_{1}^{2}}{\kappa}G_{k}\left(0\right)+\frac{a_{2}^{2}}{\kappa}G_{k}\left(0\right)+2\frac{a_{1}a_{2}}{\kappa}G_{k}\left(x_{1}-x_{2}\right)\right]
\end{eqnarray*}
and obtain the distance dependence by evaluating the following manifestly well
defined ratio for arbitrary $x_{1}\neq x_{2}$ :
\begin{eqnarray}
\frac{\langle e^{2a_{1}\chi\left(x_{1}\right)}e^{2a_{2}\chi\left(x_{2}\right)}\rangle_{k}}{\langle e^{2a_{1}\chi\left(x_{1}\right)}\rangle_{\mu}\langle e^{2a_{2}\chi\left(x_{2}\right)}\rangle_{\mu}} & = & \exp\left[\frac{a_{1}^{2}+a_{2}^{2}}{\kappa}\left(G_{k}\left(0\right)-G_{\mu}\left(0\right)\right)+2\frac{a_{1}a_{2}}{\kappa}G_{k}\left(x_{1}-x_{2}\right)\right]\nonumber \\
 & = & \left(\frac{k}{\mu}\right)^{-\frac{1}{2\pi}\frac{a_{1}^{2}}{\kappa}}\left(\frac{k}{\mu}\right)^{-\frac{1}{2\pi}\frac{a_{2}^{2}}{\kappa}}\left(k|x_{1}-x_{2}|\right)^{-\frac{a_{1}a_{2}}{\kappa\pi}}\label{eq:+++-martin-corrections}\\
 & = & \left(\frac{k}{\mu}\right)^{-\frac{1}{2\pi}\frac{1}{\kappa}\left(a_{1}+a_{2}\right)^{2}}\left(\mu|x_{1}-x_{2}|\right)^{-\frac{a_{1}a_{2}}{\kappa\pi}}\,.\nonumber 
\end{eqnarray}
From (\ref{eq:+++-martin-corrections}) it is clear that the IR limit $k\rightarrow0$
can be meaningfully taken only if $a_{1}=-a_{2}$. This condition
of charge neutrality is a well known feature of such Coulomb
gas calculations, see for instance \cite{ID89}. 

The problem of the $k\rightarrow 0$ limit presents itself differently depending on the sign of $\kappa$. If
$\kappa>0$ the correlator (\ref{eq:+++-martin-corrections}) with
$a_{1}\neq-a_{2}$ diverges for $k\rightarrow0$ at fixed $\mu\neq0$,
while it vanishes in this limit when $\kappa<0$. In the usual Coulomb
gas interpretation \cite{M10} this leads to the requirement of
charge neutrality, $\sum a_{i}=1$, which we shall not discuss further
here since our emphasis was on showing how the FRGE for composite
operators relates to the standard methods. 

Let us note that, imposing the condition of charge neutrality $a_1 = -a_2$, one can obtain the power law in (\ref{eq:+++-martin-corrections}),
i.e.~the distance dependence $\propto \left| x_1 -x_2\right|^{\frac{a_1 a_2}{\kappa \pi}} = \left| x_1 -x_2\right|^{\frac{a_1^2}{\kappa \pi}}$,
by assuming that the correlation
function of two composite operators $O_1$ and $O_2$ is determined by their individual (anomalous)
scale dimension. 
In the case at hand it is given
by $-\gamma_{O_{1}}-\gamma_{O_{2}}=\frac{a^{2}_1}{2\kappa\pi}+\frac{a^{2}_1}{2\kappa\pi}=\frac{a^{2}}{\kappa\pi}$.
This yields the same power law behaviour as in (\ref{eq:+++-martin-corrections})
if charge neutrality condition holds.

Let us also note that we can use (\ref{eq:34-one-point-martin-corrections})
in order to eliminate the normalization point $\mu$ from (\ref{eq:+++-martin-corrections})
yielding
\begin{eqnarray*}
\langle e^{2a_{1}\chi\left(x_{1}\right)}e^{2a_{2}\chi\left(x_{2}\right)}\rangle_{k} & = & \left(k\left|x_{1}-x_{2}\right|\right)^{-\frac{a_{1}a_{2}}{\kappa\pi}}\langle e^{2a_{1}\chi\left(x_{1}\right)}\rangle_{k}\langle e^{2a_{2}\chi\left(x_{2}\right)}\rangle_{k}\,.
\end{eqnarray*}
This relates the product of two expectation values of normal
ordered exponentials to the expectation value of their product.

\subsection{Scaling of geodesics and of generic curves}

Let us consider the following functional depending on the metric alone:
\begin{equation}
L_{g}\equiv L\left[x_{g}\left(\cdot\right);g\right]=\int_{0}^{1}ds\,\sqrt{g_{\mu\nu}\left(x_{g}\left(s\right)\right)\dot{x}_{g}^{\mu}\left(s\right)\dot{x}_{g}^{\nu}\left(s\right)}\,.\label{eq:34-prime-geodesics-martin-corrections}
\end{equation}
Here $x_{g}^{\mu}\left(s\right)$ parametrizes the {\it geodesic} determined
by $g_{\mu\nu}$ and connecting $x_{g}\left(0\right)$
and $x_{g}\left(1\right)$. 

Hence, in comparison with (\ref{eq:27-prime-martin-corrections})
the length (\ref{eq:34-prime-geodesics-martin-corrections}) has an
additional source of metric dependence via the curve considered. As
a result, the functional $L_{g}$ is an even more complicated composite
operator built from the quantum metric. We will compute its 
anomalous dimension which describes how the quantum average of $L_{g}$
responds to a scale variation. 

Let us stress that since $x_{g}\left(s\right)$ occuring in $L_{g}$
solves the geodesics equation, and thus depends implicitly on the
metric, a variation of $g_{\mu\nu}$ changes the geodesics equation
and its associated solution. This is a crucial difference for our
computation since it renders the Hessians of the composite operators
$L_{g}$ and $L$ different. Based on this observation we expect that the
scaling dimensions of $L_{g}$ and $L$ may turn out different. In general, if one wishes
to recover the result regarding $L$,
for generic fixed curves, not necessarily geodesics,
one just needs to drop from
the Hessian of $L_{g}$ the extra contribution which do not appear
in the Hessian of $L$.

Let us interpret $g_{\mu\nu}\equiv g_{\mu\nu}^{\rm bare}$ in (\ref{eq:34-prime-geodesics-martin-corrections})
as the bare metric now, and let us parametrize it as $g_{\mu\nu}=e^{2\chi}\delta_{\mu\nu}$
so that
\begin{eqnarray*}
L_{g} & = & \int_{0}^{1}ds\,\left|\dot{x}_{g}^{\mu}\left(s\right)\right|e^{\chi\left(x_{g}\left(s\right)\right)}\,.
\end{eqnarray*}
The explicit form of $L_{g}$ can be worked out explicitly in two dimensions, expanding
with respect to $\chi$ \cite{David:1991es,Braune:1994mk}. Up to the second order in $\chi$
it reads
\begin{eqnarray}
L_{g} & = & \left|x_{0}-y_{0}\right|\int_{0}^{1}ds\,\left[1+\chi\left(x\left(s\right)\right)+\frac{1}{2}\chi\left(x\left(s\right)\right)^{2}\right]\nonumber \\
 &  & -\frac{1}{2}\left|x_{0}-y_{0}\right|^{3}\int_{0}^{1}du\int_{0}^{1}dv\,\partial_{\bot}\chi\left(x\left(u\right)\right)D_{u,v}\partial_{\bot}\chi\left(x\left(v\right)\right)\,,\label{eq:length_geodesic_2nd_order_phi}
\end{eqnarray}
where $\left|x_{0}-y_{0}\right|$ is the flat spacetime distance between
the two points connected by the geodesics, and
\begin{eqnarray*}
x^{\mu}\left(s\right) & \equiv & x_{0}^{\mu}\left(1-s\right)+y_{0}^{\mu}s\\
\partial_{\bot}\chi & \equiv & \varepsilon_{\;\nu}^{\mu}\frac{y_{0}^{\nu}-x_{0}^{\nu}}{\left|x_{0}-y_{0}\right|}\partial_{\mu}\phi\\
D_{u,v} & \equiv & v\left(1-u\right)\theta\left(u-v\right)+u\left(1-v\right)\theta\left(v-u\right)\,.
\end{eqnarray*}

We shall read off the anomalous dimension of $L_{g}$, which we denote
$\gamma_{L_{g}}$, from equation (\ref{eq:flow_eq_gamma_Z_no_mixing})
by projecting on the monomial $L_{g}$ in flat spacetime, that is,
by setting $\chi=0$ after having computed the Hessian of $L_g$. 
This means that we have to single out the terms
proportional to $\left|x_{0}-y_{0}\right|$ when we compute the trace
on RHS of (\ref{eq:flow_eq_gamma_Z_no_mixing}). In this manner, equation
(\ref{eq:flow_eq_gamma_Z_no_mixing}) will read
\begin{eqnarray}
\gamma_{L_{g}}\left|x_{0}-y_{0}\right| & = & -\frac{1}{2}\mbox{Tr}\left[{\cal G}_{k}\cdot L_{g}^{\left(2\right)}\cdot{\cal G}_{k}\cdot\partial_{t}{\cal R}_{k}\right]\label{eq:flow_anomalous_dim_lg}
\end{eqnarray}
where ${\cal G}_{k}$ is the regularized inverse propagator ${\cal G}_{k}=\left(\Gamma_{k}^{\left(2\right)}+{\cal R}_{k}\right)^{-1}$
and $L_{g}^{\left(2\right)}$ is the Hessian of the geodesic length
operator. Let us observe that furthermore
\begin{eqnarray}
\gamma_{L_{g}}\left|x_{0}-y_{0}\right| & = & -\frac{1}{2}\int \frac{d^2p}{\left( 2\pi \right)^2}\, \langle p|{\cal G}_{k}\cdot L_{g}^{\left(2\right)}\cdot{\cal G}_{k}\cdot\partial_{t}{\cal R}_{k}|p\rangle\nonumber \\
 & = & -\frac{1}{2}\int \frac{d^2p}{\left( 2\pi \right)^2}\, {\cal G}_{k}\left(p\right){\cal G}_{k}\left(p\right)\partial_{t}{\cal R}_{k}\left(p\right)\langle p|L_{g}^{\left(2\right)}|p\rangle\,,\label{eq:25-prime-martin-corrections}
\end{eqnarray}
where we expressed, in flat spacetime, the trace as a single momentum integral.
As a result, we are left with finding the explicit form of the matrix
element $\langle p|L_{g}^{\left(2\right)}|p\rangle$. 
It can be obtained as follows. 
After a Fourier transform, $\langle p|L_{g}^{\left(2\right)}|p\rangle$ is seen to consist of three
pieces, labelled $a$, $b$ and $c$, respectively:
\begin{eqnarray}
\langle p|L_{g}^{\left(2\right)}|p\rangle & = & \int d^2x^{\prime} d^2x^{\prime\prime}\, 
e^{ip\cdot\left(x^{\prime}-x^{\prime\prime}\right)}\langle x^{\prime}|L_{g}^{\left(2\right)}|x^{\prime\prime}\rangle\equiv{\cal P}_{a}+{\cal P}_{b}+{\cal P}_{c}\,.\label{eq:Hessian_lg_from_space_to_Fourier}
\end{eqnarray}
The matrix element $\langle x^{\prime}|L_{g}^{\left(2\right)}|x^{\prime\prime}\rangle\equiv{\cal Q}_{a}+{\cal Q}_{b}+{\cal Q}_{c}$,
i.e.~the Hessian $\delta^{2}L_{g}/\delta\phi\left(x^{\prime}\right)\delta\phi\left(x^{\prime\prime}\right)$
of (\ref{eq:length_geodesic_2nd_order_phi}), consists of the following
three terms:
\begin{eqnarray}
{\cal Q}_{a} & \equiv & \left|x_{0}-y_{0}\right|\int_{0}^{1}ds\,\delta\left(x^{\prime}-x\left(s\right)\right)\delta\left(x^{\prime\prime}-x\left(s\right)\right)\nonumber \\
{\cal Q}_{b} & \equiv & -\frac{1}{2}\left|x_{0}-y_{0}\right|^{3}\int_{0}^{1}du\int_{0}^{1}dv\,\partial_{\bot}\delta\left(x^{\prime}-x\left(u\right)\right)D_{u,v}\partial_{\bot}\delta\left(x^{\prime\prime}-x\left(v\right)\right)\quad\,\,\,\label{eq:Hessian_lg_real_space}\\
{\cal Q}_{c} & \equiv & -\frac{1}{2}\left|x_{0}-y_{0}\right|^{3}\int_{0}^{1}du\int_{0}^{1}dv\,\partial_{\bot}\delta\left(x^{\prime\prime}-x\left(u\right)\right)D_{u,v}\partial_{\bot}\delta\left(x^{\prime}-x\left(v\right)\right)\,.\nonumber 
\end{eqnarray}
\\
{\bf (a)}
The first term, ${\cal Q}_{a}$, in equation (\ref{eq:Hessian_lg_real_space})
is the part of the Hessian which coincides with the one present when
we compute the Hessian of the length for a {\it generic} curve, not necessarily
a geodesic. Using equations (\ref{eq:25-prime-martin-corrections})
and (\ref{eq:Hessian_lg_from_space_to_Fourier}) we see that it gives
rise to the following contribution to the RHS of (\ref{eq:25-prime-martin-corrections}):
\begin{equation}
-\frac{1}{2\pi}\left(\frac{1}{4\kappa}\left|x_{0}-y_{0}\right|\right)\,.
\end{equation}
\\
{\bf (b)}
Now we evaluate the contribution due to ${\cal Q}_{b}$. Denoting
$\xi^{\mu}\equiv\varepsilon_{\;\nu}^{\mu}\left(y_{0}^{\nu}-x_{0}^{\nu}\right)$,
it reads
\begin{equation}
{\cal Q}_{b}=-\frac{1}{2}\left|x_{0}-y_{0}\right|\int_{0}^{1}du 
\int_{0}^{1}dv\,\xi^{\mu}\frac{\partial}{\partial x^{\prime\mu}}\delta\left(x^{\prime}-x\left(u\right)\right)D_{u,v}\xi^{\nu}\frac{\partial}{\partial x^{\prime\prime\nu}}\delta\left(x^{\prime\prime}-x\left(v\right)\right)\,.
\end{equation}
Inserting this expression in equation (\ref{eq:Hessian_lg_from_space_to_Fourier})
one obtains
\begin{eqnarray}
{\cal P}_{b}=
-\frac{1}{2}\left|x_{0}-y_{0}\right|\int dx^{\prime} dx^{\prime\prime}\,
e^{ip\cdot\left(x^{\prime}-x^{\prime\prime}\right)}\int_{0}^{1}du 
\hspace{60mm}
\nonumber \\
\times \int_{0}^{1}dv\,\xi^{\mu}\left(-ip_{\mu}\right)\delta\left(x^{\prime}-x\left(u\right)\right)D_{u,v}\xi^{\nu}\left(ip_{\nu}\right)\delta\left(x^{\prime\prime}-x\left(v\right)\right) \,.
\end{eqnarray}
Integrating over $x^{\prime}$ and $x^{\prime\prime}$ we obtain the
following contribution to (\ref{eq:Hessian_lg_from_space_to_Fourier})
\begin{eqnarray}
{\cal P}_{b} & = & -\frac{1}{8}\left|x_{0}-y_{0}\right|\int_{0}^{1}du\int_{0}^{1}dv\,e^{ip\cdot\left(x\left(u\right)-x\left(v\right)\right)}D_{u,v}\xi^{\mu}\xi^{\nu}p_{\mu}p_{\nu}\,.\label{eq:+-geodesic-martin-corrections}
\end{eqnarray}
\\
{\bf (c)}
The third piece ${\cal Q}_{c}$ in equation (\ref{eq:Hessian_lg_real_space})
gives a result identical to (\ref{eq:+-geodesic-martin-corrections}) but
with $x\left(u\right)$ and $x\left(v\right)$ interchanged. Summing
the two contributions we find
\begin{eqnarray}
{\cal P}_{b}+{\cal P}_{c}  =  -\frac{1}{8}\left|x_{0}-y_{0}\right|\int_{0}^{1}du\int_{0}^{1}dv\,\left(e^{ip\cdot\left(x\left(u\right)-x\left(v\right)\right)}+e^{-ip\cdot\left(x\left(u\right)-x\left(v\right)\right)}\right)D_{u,v}\xi^{\mu}\xi^{\nu}p_{\mu}p_{\nu}\,. \,\,\,
\label{eq:++-geodesic-martin-correction}
\end{eqnarray}
Noting that $x^{\mu}\left(u\right)-x^{\mu}\left(v\right)=-\left(u-v\right)\left(x_{0}^{\mu}-y_{0}^{\mu}\right)$,
the integral over the parameters $u$ and $v$ can be performed now, and
we have
\begin{eqnarray}
\int_{0}^{1}du\int_{0}^{1}dv\,\left(e^{ip\cdot\left(x\left(u\right)-x\left(v\right)\right)}+e^{-ip\cdot\left(x\left(u\right)-x\left(v\right)\right)}\right)D_{u,v} 
\hspace{5cm} \nonumber \\ 
=  2\int_{0}^{1}du\int_{0}^{1}dv\,\cos\left(p\cdot\left(x\left(u\right)-x\left(v\right)\right)\right)D_{u,v}
\hspace{8mm} \nonumber \\
= 2\frac{\left(-2+2\cos\left(p\cdot\left(x_{0}-y_{0}\right)\right)+\left(p\cdot\left(x_{0}-y_{0}\right)\right)^{2}\right)}{\left(p\cdot\left(x_{0}-y_{0}\right)\right)^{4}}\,.
\end{eqnarray}
This expression shows that the RG flow generates infinitely many monomials proportional
to $\left|x_{0}-y_{0}\right|^{n}$ on the RHS of the flow
equation for the composite operator. To make them explicit we expand the above
term as a power series in $p\cdot\left(x_{0}-y_{0}\right)$, finding
\begin{eqnarray*}
2\frac{-2+2\cos\Bigr(p\cdot\left(x_{0}-y_{0}\right)\Bigr)+\Bigr(p\cdot\left(x_{0}-y_{0}\right)\Bigr)^{2}}{\Bigr(p\cdot\left(x_{0}-y_{0}\right)\Bigr)^{4}} & = & \frac{1}{6}-\frac{1}{180}\Bigr(p\cdot\left(x_{0}-y_{0}\right)\Bigr)^{2}+\cdots\,.
\end{eqnarray*}
So we obtain for (\ref{eq:++-geodesic-martin-correction}) finally
\begin{eqnarray}
{\cal P}_{b}+{\cal P}_{c} =  -\frac{1}{2}\left|x_{0}-y_{0}\right|\left(\frac{1}{6}-\frac{1}{180}\left(p\cdot\left(x_{0}-y_{0}\right)\right)^{2}+\cdots\right)\xi^{\mu}\xi^{\nu}p_{\mu}p_{\nu}\,. \label{eq:52prime-martin-corrections}
\end{eqnarray}

Under our approximation we must consistently neglect all monomials with $\left(x_{0}-y_{0}\right)$-dependencies
different from $\left|x_{0}-y_{0}\right|$ when evaluating the RHS
of equation (\ref{eq:flow_anomalous_dim_lg}). 
As a consequence, equation (\ref{eq:52prime-martin-corrections})
implies that {\it ${\cal P}_{b}+{\cal P}_{c}$ gives no contribution
to the anomalous dimension $\gamma_{L_{g}}$}. 

This is seen easily after
a symmetric integration over the momenta in (\ref{eq:25-prime-martin-corrections}) with (\ref{eq:52prime-martin-corrections}).
It replaces $p_{\mu}p_{\nu}\rightarrow p^{2}\delta_{\mu\nu}/2$ in the leading term,
so that effectively 
\begin{eqnarray*}
{\cal P}_{b}+{\cal P}_{c} & = & -\frac{1}{2}\left|x_{0}-y_{0}\right|\left(\frac{1}{6}+O\left(\left(x_{0}-y_{0}\right)^{2}\right)\right)\xi^{\mu}\xi^{\nu}p^{2}\frac{\delta_{\mu\nu}}{2}\\
 & = & -\frac{1}{2}\left|x_{0}-y_{0}\right|\left(\frac{1}{6}+O\left(\left(x_{0}-y_{0}\right)^{2}\right)\right)\xi^{2}\frac{p^{2}}{2}\,.
\end{eqnarray*}
We also observe that
$
\xi^{2}=\varepsilon_{\;\rho}^{\mu}\left(y_{0}^{\rho}-x_{0}^{\rho}\right) \delta_{\mu\nu} 
\varepsilon_{\;\sigma}^{\nu}\left(y_{0}^{\sigma}-x_{0}^{\sigma}\right)=\left|x_{0}-y_{0}\right|^{2}\,,
$
since $\varepsilon_{\;\rho}^{\mu}\varepsilon_{\mu\sigma}=\delta_{\rho\sigma}$.
This then implies that in (\ref{eq:52prime-martin-corrections})
the lowest order in $\left|x_{0}-y_{0}\right|$ is proportional to
$\left|x_{0}-y_{0}\right|^{3}$:
${\cal P}_{b}+{\cal P}_{c}=-\frac{1}{4}\left|x_{0}-y_{0}\right|^{3}+\cdots$. 
As a result, ${\cal P}_{b}+{\cal P}_{c}$
contains no term that matches the linear one on the LHS of (\ref{eq:25-prime-martin-corrections})
and could contribute to $\gamma_{L_{g}}$. 

Hence our final conclusion is that the anomalous dimension $\gamma_{L_{g}}$
is determined solely by the contribution coming from ${\cal P}_{a}$.
It reads
\begin{equation}
 \fbox{$\displaystyle
\gamma_{L_{g}}  =  -\frac{1}{2\pi}\frac{1}{4\kappa}\,.\label{eq:anomalous-dim-geod-length-2D}
  $}
 \end{equation}
In turn this demonstrates that 
{\it the anomalous dimensions for the length of a geodesics, $\gamma_{L_{g}}$,
and for a generic curve, respectively, are equal within
the approximation employed}. 

However, let us stress that in general
they are likely to be different once the mixing in the running
of the geodesic length operator is taken into account. 
Nevertheless, one may interpret our result as an indication that the anomalous dimensions
of the length of a generic curve and of a geodesic are not too different
(at least in two dimensions). 

We emphasize that the geodesic length enters in many potentially observable correlation
functions. 
For instance, given two local operators $O_1$ and $O_2$ it would be interesting to compute\cite{Ambjorn:1997di,Hamber:2009zz}
\begin{eqnarray*}
G\left(r\right) & \equiv & \langle\int d^{d}x\sqrt{g\left(x\right)}\int d^{d}y\sqrt{g\left(y\right)}\,O_{1}\left(x\right)O_{2}\left(y\right)\delta\left(r-L_{g}\left(x,y\right)\right)\rangle\,.
\end{eqnarray*}
Clearly, $L_{g}$ being a non-trivial composite operator, the scaling
analysis of $G\left(r\right)$ is affected by the presence of the
delta function involving $L_{g}$. Once the full scaling dimensions
of the operators involved are known, it is straightforward to invoke
scaling arguments for this type of correlation functions, see for
instance \cite{Codello:2014wfa}.

\section{Conclusions and outlook}

In this work we considered the role of composite operators in the Asymptotic
Safety program. 
We have argued that the introduction of composite operators
via suitable sources is convenient in a number of cases. In particular,
our framework makes it possible to consider geometrical objects, like
the length of an arbitrary curve or of a geodesic, whose quantum properties would hardly be seen
in any realistic truncation for the EAA. 
Moreover we demonstrated that particular
operators, like a composite metric in the vielbein formalism, require
a careful regularization and renormalization procedure,
on top of that related to the EAA, to be meaningfully defined. 
Within the FRG setting we systematized this procedure for arbitrary composite operators.
In general, the introduction of composite operators is useful whenever
one wishes to investigate the quantum properties of operators
that are not contained in the (exact) EAA, or in the truncation considered. 

In sections \ref{sec:Floating-normalization-point-and-FRG} and \ref{sub:Composite-operators-in-FRG}
we have reviewed the inclusion of composite operators in the EAA
formalism and discussed a method which allows to identify the scaling
properties of the composite operators at the fixed point. In section
\ref{sec:Composite-metrics-in-the-CREH} we considered the CREH truncation
and studied the case of composite metrics in this setting. 
The CREH example made it explicit that a dedicated regularization and subsequent (re-)normalization 
is necessary in order to define the metric whenever the latter is a composite field. 
As such, the composite metric analyzed in section \ref{sec:Composite-metrics-in-the-CREH}
can be viewed as a toy model for
the composite metric in the vielbein formalism. 

The CREH model also illustrates nicely that by choosing different field parametrizations quantum
corrections can be changed crucially both in the EAA and the composite operator.
As an extreme example, with $g_{\mu\nu} =\phi^2 \delta_{\mu\nu}$ the metric is a composite operator whereas with 
the the alternative parametrization $g_{\mu\nu} =\psi \delta_{\mu\nu}$ is not. 
But in the latter case also the CREH ansatz (\ref{eq:CREH_truncation_generic_dimension})
acquires a different form, the kinetic term $\propto \left(\partial_\mu \psi \right)^2/\psi $ is no 
longer bilinear in the dynamical field, and so
that the running of the couplings involved will be different.

In section \ref{sec:Volume-operators-in-AS}
we tested further our framework by computing the anomalous dimensions of the
volume and the length operators. Finally, in section \ref{sec:Two-dimensional-QG},
we considered Liouville theory in two dimensions. In particular, we
computed the correlation functions of composite metrics and the anomalous
dimension of the geodesic distance. 

In general, since in our computations the approximations
and ans{\"a}tze were too simple still, we do not expect our results
to be quantitatively precise. However, it is important to note that
the framework introduced in this work has allowed for the first time
to give an estimate of the quantum properties of geometrical objects,
like the length of a curve, that have never been considered before for the case
of asymptotically safe quantum gravity.

We would like to remark that the final purpose of explicitly keeping track of 
selected composite operators
is making contact with quantum gravitational observables. Clearly,
this is beyond the scope of the present paper but we made a first
step towards this goal. Indeed, as we argued in the introduction,
in order to consider certain types of observables it is unavoidable
to introduce further operators on top of those present in the gravitational EAA. In
this sense, it would be natural to follow the logic of the two dimensional
case where fixed-volume and fixed-geodesic distance functionals have
been discussed in detail. Ultimately, indeed, we believe that a comparison
between the different attempts to define a well defined gravitational
path integral can only be made by considering observable quantities. 

Summarizing, we
believe that the framework developed in this work opens the door to
new avenues in comparing different approaches to quantum gravity and
gives a viable road to access observables in the Asymptotic Safety
scenario for quantum gravity.


\clearpage
\appendix


\section{Liouville theory: KPZ scaling \label{sub:KPZ-scaling}}

Scaling arguments in quantum gravity have been particularly fruitful
in two dimensions. Here we shall derive the so called KPZ relations \cite{KPZ88,DDK88,Chamseddine:1988tu}
following the notation adopted in \cite{Wa93}. The partition function
can be written as follows:
\begin{equation}
Z =  \int{\cal D}\phi\exp\left[-\left(\frac{25-c}{48\pi}\right) \int d^2x \sqrt{g}
\left(\frac{1}{2}g^{\mu\nu}\partial_\mu\phi \partial_\nu \phi+R\phi\right)\right]\,. \label{eq:A1-martin-corrections}
\end{equation}
Now, the bare action in the exponent of (\ref{eq:A1-martin-corrections}) 
enjoys the trivial symmetry $g_{\mu\nu}\rightarrow e^{\sigma \left(x \right)}g_{\mu\nu},\phi\rightarrow\phi-\sigma\left(x \right)$
since it can be rewritten as a functional of the product $e^{\phi}g_{\mu\nu}$ only.
This means that the bare (``classical'') action is annihilated when one acts
on it with the operator
\begin{eqnarray*}
{\cal L} & \equiv & \left(g_{\mu\nu} \left(x \right) \frac{\delta}{\delta g_{\mu\nu}\left(x \right) }-\frac{\delta}{\delta\phi \left(x \right) }\right)\,.
\end{eqnarray*}
We require that this invariance is enjoyed also at the
quantum level by observables. In particular we shall construct the
diffeomorphism invariant operator
\begin{eqnarray*}
O & \equiv & \int d^{2}x\sqrt{g}\,e^{\alpha\phi}\,.
\end{eqnarray*}
One notices that for $\alpha\neq1$ the classical functional $O\equiv O\left[\phi; g \right]$ 
is not invariant under the $\sigma$-transformation.
The reason for considering a general parameter $\alpha$
is that we will determine its value such that 
\begin{eqnarray}
{\cal L}\langle O\rangle & = & 0\,,\label{eq:trivial_weyl_invariance_on_O}
\end{eqnarray}
which is to say that the operator ${\cal L}$ annihilates $O$ at the
quantum level. Indeed, quantum corrections to the naive scaling properties
will force us to fix $\alpha$ to some specific value different from unity.

Let us turn to the scaling properties of the operator $e^{\alpha\phi \left(x \right)}$.
At the quantum level, $e^{\alpha\phi \left(x \right)}$ will acquire an anomalous dimension
$\gamma$ which we will compute later on. In the fixed point regime,
the anomalous dimension enters the corresponding Callan-Symanzik equation as
follows
\begin{eqnarray}
\left(\mu\partial_{\mu}+\gamma\right)\langle e^{\alpha\phi}\rangle  =  0\,. \label{eq:54prime-martin-corrections}
\end{eqnarray}
As usual, to deduce the scaling properties of $\langle e^{\alpha\phi}\rangle$
we need to eliminate the $\mu$-derivative from the equation (\ref{eq:54prime-martin-corrections}). This can be done by
means of simple dimensional analysis. When working in curved spacetime
one may choose either the coordinates or the metric to be dimensionful.
In our case it is natural to take the coordinates dimensionless as
they are merely variables devoid of any particular meaning. Moreover
we note that $e^{\alpha\phi}$ is classically dimensionless.
Then dimensional analysis implies
\begin{eqnarray}
\left(\mu\partial_{\mu}-2g_{\mu\nu}\frac{\delta}{\delta g_{\mu\nu}}\right)\langle e^{\alpha\phi}\rangle  =  0\,. 
\label{eq:54second-martin-corrections}
\end{eqnarray}
Eliminating the $\mu\partial_\mu$ term from equations (\ref{eq:54prime-martin-corrections}) and (\ref{eq:54second-martin-corrections}) yields:
\begin{eqnarray}
\left(2g_{\mu\nu}\frac{\delta}{\delta g_{\mu\nu}}+\gamma\right)\langle e^{\alpha\phi}\rangle & = & 0\,.\label{eq:variazione_g_exp_alpha_phi}
\end{eqnarray}

Now let us determine the action of ${\cal L}$ on $\langle O\rangle$:
\begin{eqnarray*}
{\cal L}\langle O\rangle & = & \left(g_{\mu\nu}\frac{\delta}{\delta g_{\mu\nu}}-\frac{\delta}{\delta\phi}\right)\langle O\rangle\\
 & = & \left(g_{\mu\nu}\frac{\delta}{\delta g_{\mu\nu}}-\frac{\delta}{\delta\phi}\right)\langle\int d^{2}x\sqrt{g}\,e^{\alpha\phi}\rangle\\
 & = & \left(g_{\mu\nu}\frac{\delta}{\delta g_{\mu\nu}}-\frac{\delta}{\delta\phi}\right)\int d^{2}x\sqrt{g}\,\langle e^{\alpha\phi}\rangle\,,
\end{eqnarray*}
where we brought the volume element outside the average $\langle\cdot\rangle$ 
since the latter is not dynamical. Now we note that the functional
derivative with respect to the metric acts obviously on $\sqrt{g}$,
but also on the implicit dependence of $\langle e^{\alpha\phi}\rangle$
on the metric. This latter dependence is easily obtained from equation
(\ref{eq:variazione_g_exp_alpha_phi}). Therefore we have the following
contributions:
\begin{eqnarray*}
g_{\mu\nu}\left(x\right)\frac{\delta}{\delta g_{\mu\nu} \left(x\right)} \sqrt{g} \left(x^\prime \right) & = & \sqrt{g}\, \delta \left(x-x^\prime \right) \\
g_{\mu\nu}\left(x\right) \frac{\delta}{\delta g_{\mu\nu}\left(x\right)}\langle e^{\alpha\phi \left(x^\prime \right)}\rangle & = & -\frac{\gamma}{2}\langle e^{\alpha\phi}\rangle \delta \left(x-x^\prime \right) \\
-\frac{\delta}{\delta\phi \left(x\right)}\langle e^{\alpha\phi \left(x^\prime \right)}\rangle & = & -\alpha\langle e^{\alpha\phi}\rangle \delta \left(x-x^\prime \right) \,.
\end{eqnarray*}
Summing all the terms we can finally write equation (\ref{eq:trivial_weyl_invariance_on_O})
as
\begin{eqnarray}
 1-\alpha-\frac{\gamma}{2} & = & 0\,.\label{eq:KPZ_gamma}
\end{eqnarray}
We shall see in a moment that this is the celebrated KPZ relation.

Finally we come to our point, the actual computation of $\gamma$. According to
the discussion in section \ref{sub:Composite-operators-in-FRG} we
need to evaluate $\gamma=Z_{e^{\alpha\phi}}^{-1}\partial_{t}Z_{e^{\alpha\phi}}$
via equation (\ref{eq:flow_eq_gamma_Z_no_mixing}). 
It is sufficient and particularly convenient
to set $g_{\mu\nu}=\delta_{\mu\nu}$. Doing so one obtains:
\begin{eqnarray*}
\left(Z_{e^{\alpha\phi}}^{-1}\partial_{t}Z_{e^{\alpha\phi}}\right)e^{\alpha\phi \left(x\right)} & = & -\frac{1}{2}
\mbox{Tr} \left[ \frac{1}{\left(\frac{25-c}{48\pi}\right)\left(-\Box+R_{k}\right)}\left(\alpha^{2}e^{\alpha\phi \left(x\right)}\right)\frac{1}{\left(\frac{25-c}{48\pi}\right)\left(-\Box+R_{k}\right)}\left(\frac{25-c}{48\pi}\right)\partial_{t}R_{k} \right] \\
 & = & -\frac{12\alpha^{2}}{25-c}\,e^{\alpha\phi \left(x\right)}\,.
\end{eqnarray*}
Inserting this value for $\gamma$ in equation (\ref{eq:KPZ_gamma})
we obtain:
\begin{equation}
\fbox{$\displaystyle
1-\alpha+\frac{6\alpha^{2}}{25-c} =  0\,.
  $}
\end{equation}
This relation is a well known result in Liouville gravity,
the basis in particular for the ``gravitational dressing'' of arbitrary matter field operators,
see e.g.~\cite{Wa93}.

We can rephrase our arguments also in the following way: The operator
$e^{\alpha\phi}$ has a vanishing classical mass dimension. 
However, quantum corrections afflict $e^{\alpha\phi}$ with an anomalous dimension
equal to $\gamma$. Under a Weyl rescaling an operator with mass dimension
$\gamma$ is transformed by an overall factor $\left(e^{-\frac{\sigma}{2}}\right)^{\gamma}$.\footnote{
We recall that in the conventions adopted in this Appendix the metric transforms 
via $g_{\mu\nu}\rightarrow e^{\sigma \left(x \right)}g_{\mu\nu}$ under a Weyl transformation.}
Therefore, at the quantum level, the operator $\sqrt{g}e^{\alpha\phi}$
gets transformed by an overall factor $e^{\sigma}e^{-\alpha\sigma}e^{-\frac{\sigma}{2}\gamma}$,
where the first exponential comes from the determinant of the metric
while the other exponentials come from the classical and quantum
scaling properties of the operator $e^{\alpha\phi}$, respectively.
We note that the exponent of the overall scaling factor $e^{\sigma \left(1-\alpha -\gamma/2 \right)}$ is precisely
the LHS of equation (\ref{eq:KPZ_gamma}) which we
require to vanish in order to satisfy the invariance under the operator
${\cal L}$ at the quantum level.

For further details on the EAA approach to Liouville theory and KPZ scaling we refer to \cite{Reuter:1996eg}.

\end{spacing}


\end{document}